\documentclass[preprint, aip, ajp]{revtex4-2}
\usepackage{graphicx}
\usepackage{amsmath}
\usepackage{hyperref}
\hypersetup{nolinks=true} 

\usepackage{siunitx}
\usepackage{booktabs}
\usepackage{esvect}

\usepackage{color, soul}

\begin{document}

\title{Investigating the Magnetic Field outside Small Accelerator Magnet Analogs via Experiment, Simulation, and Theory}

\author{Kelley D.\ Sullivan}

\author{Antara Sen}

\author{M.\ C.\ Sullivan}
\email{mcsullivan@ithaca.edu}

\affiliation{Department of Physics \& Astronomy, Ithaca College, Ithaca, New York 14850}

\date{\today}

\begin{abstract}

Particle accelerators use powerful and complex magnetic fields to turn, shape, and eventually collide beams of near-light-speed particles, yet the fundamental magnetic principles behind the accelerator magnets can be understood by undergraduate students.  In this paper we use small-scale accelerator magnet analogs in a multi-faceted, low-cost exploration of the magnetic field exterior to accelerator magnets.  These fields are best understood using the multipole expansion of the field.  If we assume that the magnetic field is created by ideal magnetic dipoles, we can derive a theoretical model that shows that each accelerator magnet configuration is dominated by a single multipole moment and obeys $B \propto 1/r^{l+2}$, where $l$ is the multipole order (with $l = 1, 2, 3, 4$ for the dipole, quadrupole, octopole, and hexadecapole moments, respectively).  Using commercially available NdFeB magnets and the magnetic field sensor inside a smartphone, we experimentally verify the power-law dependence of the accelerator magnet configurations.  Finally, we use the open-source Python library Magpylib to simulate the magnetic field of the permanent magnet configurations, showing good agreement between theory, experiment, and simulation.

\end{abstract}

\maketitle

\section{Introduction}
The advanced laboratory project described in this article uses accelerator magnets as the motivation -- the ``hook" -- but the learning goals of the project are mathematical derivation, numerical computation, visualization via simulation, experiment design and execution, and data collection and analysis. This project can be scaled to fit the learning objectives of intermediate or advanced laboratory courses and can be modified to match the desired skill development for an individual student. Requiring, at minimum, only a handful of magnets, 3D-printed magnet holders, a few raw materials typically found around the home, a smartphone, and some ingenuity, this project can be completed at low cost and can be conducted in-person or remotely.

\begin{figure}
   \centering
   \includegraphics[width=0.85\linewidth]{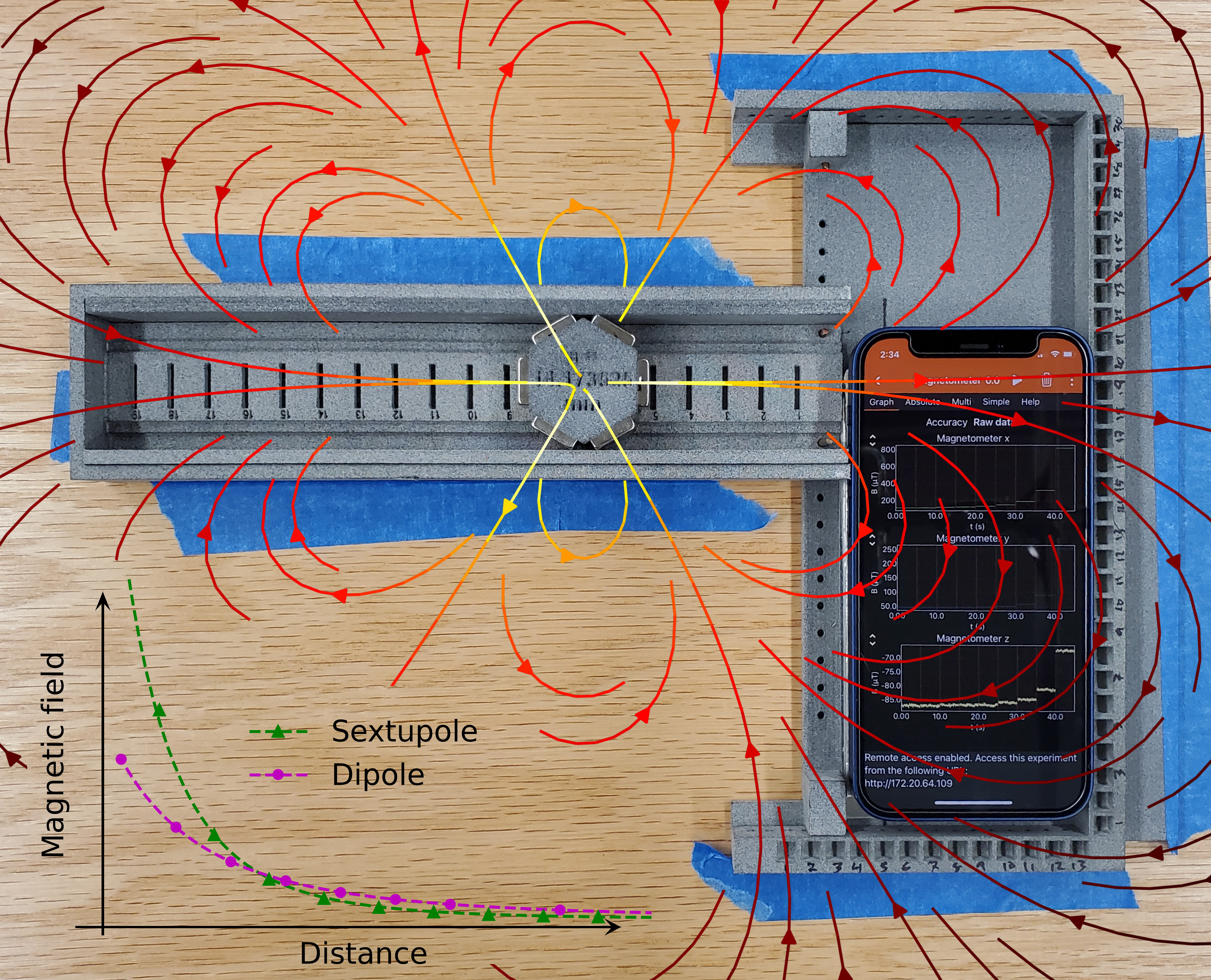}
   \caption{A visual integration of the project's many facets. The 3D-printed smartphone bed (positioned vertically to the right) securely holds the smartphone.  The 3D-printed extension arm (positioned horizontally across the center of the figure) contains the sextupole configuration of magnets. Magnetic field lines (from Magpylib simulation) are shown centered on the magnet configuration. The magnetic field sensor is located in the upper left corner of the smartphone. The inset shows experimental data (markers) and theory curves (dashed lines) for the sextupole ($l=3$) and dipole ($l=1$) magnet configurations, verifying the expected $1/r^{l+2}$ behavior of the magnetic field strength as a function of radial distance from the center of the configurations.}
   \label{fig:overview}
\end{figure}

Particle accelerators are well known for their impressive and large mechanical systems, including magnetic focusing and guiding systems.\cite{niceaccelerator}  The particle beam usually travels through the center of complex magnet configurations that surround the beam with the poles of electromagnets. These magnet configurations have been designed to bend (dipole) or focus (quadrupole) the beam, or to correct chromatic aberrations (sextupole)\cite{niceaccelerator, CERN_JAIcourse} -- several even have their own Wikipedia pages.\cite{wikipedia_dipole, wikipedia_quad, wikipedia_sext}  
We present an advanced laboratory experiment that investigates accelerator magnets through inexpensive small-scale permanent magnet analogs, built by placing permanent magnets with alternating polarities on the faces of a regular polygon with an even number of sides. Fig.~\ref{fig:overview} shows a sextupole magnet configuration.  The fields inside the magnet configurations are well known\cite{niceaccelerator} (sometimes called ``contact fields," a term borrowed from atomic physics).\cite{gray09, gray10}  However, the fields \textit{outside} the configurations are not commonly studied.  Given the small size of our accelerator magnet analogs, they are perfect for use in examining these exterior fields.


In our advanced laboratory experiment, we use a smartphone to measure the magnetic field. 
The advent of smartphones has made a whole suite of physical sensors accessible to a wide audience. Smartphones have turned into sophisticated pieces of equipment\cite{smartphone_resource, elearningAJP} whose popularity rose dramatically during the pandemic as measurement devices for the undergraduate physics laboratory.  The smartphone magnetic field sensor, intended for use as a compass and in navigation and GPS, has been utilized in experiments where magnets have been used as timing sensors in pendula,\cite{manypapers} springs,\cite{manypapers} and rotations.\cite{RotationswithBTPT} The sensor, a 3-axis Hall sensor, has been used to explore the magnetic field generated by the Earth in both a laboratory context\cite{MeasEarthBEJP} and in semester-long examinations of the local magnetic field.\cite{manypapers,BfieldsinclassEJP}  The magnetic field sensor has also been used to measure the magnetic fields from wires,\cite{manypapers, simplemeasTPT} from electric rails,\cite{manypapers} from Helmholtz coils (when paired with the accelerometer),\cite{manypapers,HelmholtzMeasTPT} and recently in measurements of a dipole and linear quadrupole.\cite{arribas2015measurement, arribas2020linear} 
Most of these smartphone laboratories are intended to engage introductory physics students. Our experiment builds on Refs.\ \onlinecite{arribas2015measurement, arribas2020linear}, 
and uses the smartphone magnetic field sensor to measure the magnetic field outside an accelerator magnet analog and then compare the experimental results to theoretical predictions and simulations.  

In Fig.~\ref{fig:overview}, the magnetic field sensor inside the phone is located approximately in the same position as the vector arrowhead visible in the top left corner of the phone. The inset shows the magnetic field strength as a function of distance from the center of the magnet configuration for dipole and sextupole magnet configurations.  The data were obtained with the phone using the app Phyphox while the magnet configuration was translated to discrete positions along the extension arm.

To better understand our measurements, we must predict what the magnetic fields outside the magnet configurations should look like.  We can use simulations to predict these magnetic fields. Advances in computing have made computation and simulations common, 
and open-source programs like Python and repositories like GitHub have made these complex computations accessible and low-cost. The Python library Magpylib\cite{magpylib2020} allows users to simulate arbitrary configurations of permanent magnets and then examine the behavior of their associated fields in both space and time. The library is straightforward to use for a student with an introductory knowledge of programming. Fig.~\ref{fig:overview} shows a simulated magnetic field generated by a sextupole magnet configuration. A color gradient indicates the changing magnitude of the field (decreasing from white to yellow to red). Students can use the simulation to visualize the fields that would be present outside the accelerator magnet analogs, providing qualitative confirmation of the expected symmetry and setting requirements and constraints on the experiment design. 

Students with an interest in theoretical physics who have a working knowledge of Taylor series expansions (intermediate level) or vector calculus (advanced level) can also create a theoretical model to predict the magnetic field produced by each magnet configuration.  If each permanent magnet is replaced with an ideal dipole, then the field can be calculated analytically. To find the magnetic field outside the magnet configurations, students can use the multipole expansion. 
The expansion of arbitrary charge and current configurations into multipole moments is a fundamental tool in physics\cite{griffiths_2017} and it embodies the deconstruction of difficult problems into smaller and more tractable ones while also yielding important physical insights. 
The power-law dependence of the multipole moments, in which the field is proportional to $1/r^{l+2}$, where the power $l+2$ increases with the multipole order $l$, means that the leading multipole moment will dominate the physical behavior of the system far from the center of the system. A system with a net electric charge behaves primarily as a point charge -- the monopole term, $l=0$.  The dipole moment ($l=1$) dominates in electric systems with no net charge, and as a result, the dipole dominates bonding in chemical, biochemical, and biological systems.\cite{chembook} Magnetic systems (which have no monopoles) also behave primarily as dipoles ($l= 1$). Accelerator magnet configurations are particularly interesting because they create fields in which moments higher than the dipole moment dominate the far-field behavior.  The inset of Fig.~\ref{fig:overview} shows both the theoretical and measured the magnetic field magnitude as a function of radial distance from the center of a sextupole and a dipole magnet configuration. The magnetic field from the dipole magnet configuration generates the familiar $1/r^3$ behavior, while the sextupole magnet leading term is the octopole moment ($l=3$) and drops off more quickly as $1/r^5$. 

This experiment requires 3D printed magnet holders,\cite{EPAPS} which can be printed in-house or using a 3D printing service.\cite{Shapeways}   This experiment also provides additional opportunities for students with an interest in engineering to design and 3D print optional experimental apparatus to aid in the accurate and precise measurement of the field. In the work presented here, we use a stationary smartphone ``bed'' and extension arm (see Fig.~\ref{fig:overview}) to carefully position the smartphone and magnet configurations relative to each other to allow for easily-repeatable measurements of the field at varying distances between the center of the magnet configuration and the sensor inside the smartphone.  Our CAD files are provided in the Supplementary Material or online.\cite{supplementary, EPAPS} 

\section{Analytical Modeling} \label{sec:analytical}
\begin{figure}
   \centering
   \includegraphics[width=0.85\linewidth]{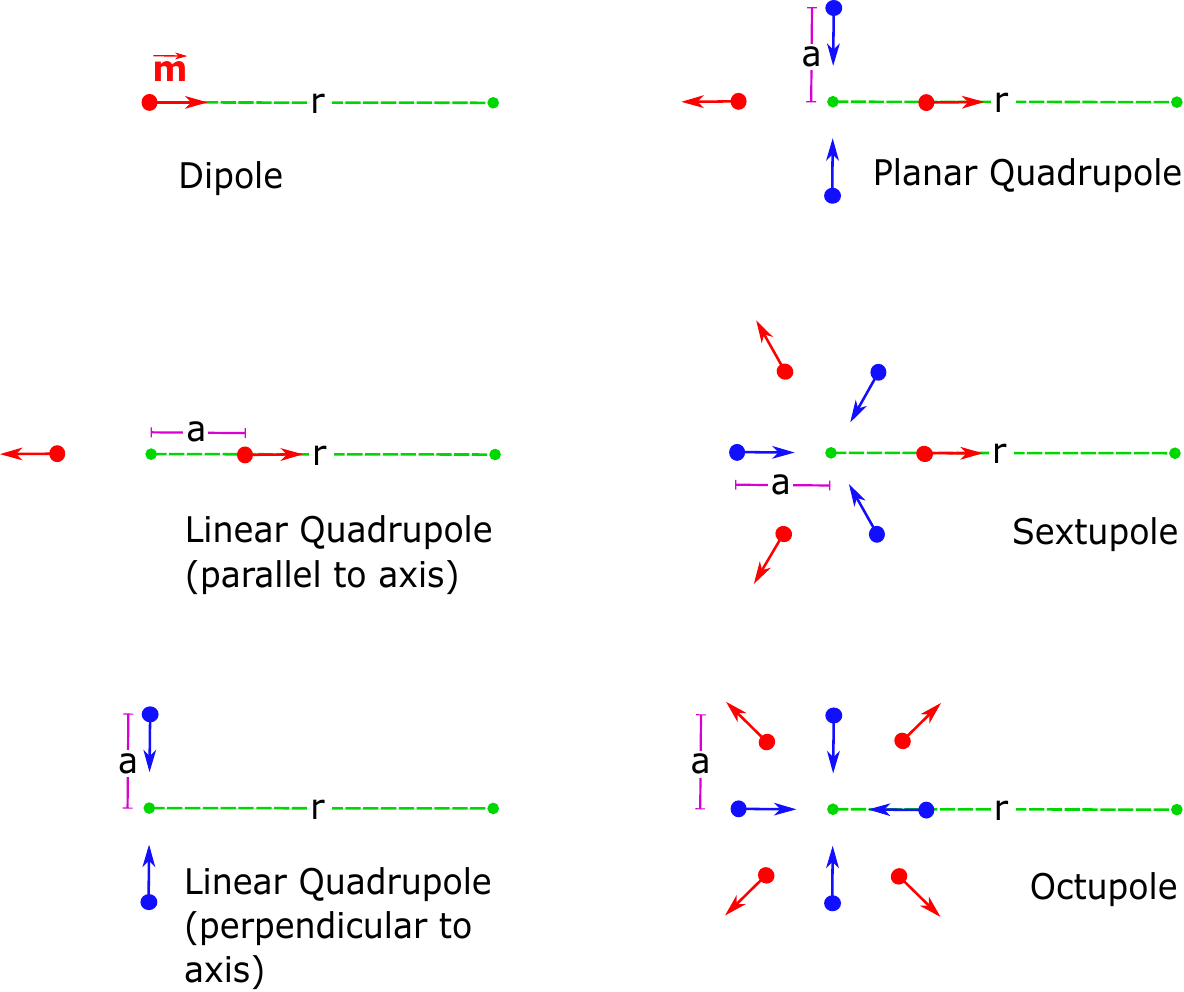}
   \caption{Arrangement of ideal dipoles with the north pole oriented either towards (blue) or away from (red) the center of the multipole configuration.  The magnetic field is evaluated at a distance $r$ from the center of the multipole configuration along an axis through one of the dipoles, except for the linear quadrupole perpendicular to the axis.  The distance from the center of the configuration to the center of each dipole is given by $a$. Multipole magnet naming convention follows the standard in accelerator physics.\cite{octupole}}
   \label{fig:multipoles from dipoles}
\end{figure}

In this article, we will present measurements of six different planar configurations of permanent magnets, represented in Fig.\ \ref{fig:multipoles from dipoles} as ideal dipoles each with dipole moment $m$.\footnote{A dipole is a current loop of area $A$ and current $I$ with dipole moment $m=IA$.  An ideal dipole is a loop of infinitesimal area $A$ and infinite current $I$ where the product $m=IA$ remains constant.\cite{griffiths_2017}}  The three simple configurations on the left have been studied or are simple extensions of previous work.\cite{griffiths_2017, arribas2020linear}  The three configurations on the right are inspired by accelerator magnets, where each dipole sits on the face of a regular polygon, with dipoles pointing alternately towards or away from the center of the configuration.  These configurations are called the quadrupole, sextupole, and octupole magnets.  In this experiment, we will consider only the field in the plane of the magnet configurations.

Unfortunately, there is inconsistent nomenclature between the multipole expansion standard in physics, chemistry, and engineering and the nomenclature used in accelerator physics.  In the expansion of fields into their multipole moments, the order $l$ gives rise to the name of the moment via the power $2^l$.  Hence, the orders $l=0, 1,2,3,4$ ($2^l = 1, 2,4,8,16$) are named the monopole, dipole, quadrupole, octopole, and hexadecapole (16-pole) moments.  In accelerator physics, it is standard to name the magnet configurations by the number of poles that face the center of the configuration (where the particle beam would travel).  Thus, in accelerator physics a dipole magnet has two poles and is used to create a constant field.\cite{wikipedia_dipole}  The quadrupole, sextupole, and octupole magnets are arranged around a square, hexagon, and octagon, respectively, as shown in Fig.\ \ref{fig:multipoles from dipoles}. To avoid confusion, we always refer to them as \textbf{moments} (e.g., the octopole moment) from the multipole expansion and \textbf{magnets} (e.g., the sextupole magnet) from accelerator physics.\cite{octupole}

To calculate the field from the magnet configurations in Fig.\ \ref{fig:multipoles from dipoles}, we should examine them in the multipole expansion.  The expansion of charge and current distributions into their multipole moments is discussed at the undergraduate level,\cite{griffiths_2017} though there is little discussion of the moments beyond the dipole.  The calculation of multipole moments from an arbitrary current distribution is well established,\cite{HardMultipoleBook} and multipoles can be calculated even from permanent magnets.\cite{wolski2019maxwells}  Analytical derivations of the magnetic field of the multipoles can be quite daunting, however, and is the subject of multiple publications in the pedagogical literature.\cite{gray78, HardMultipoleAJP, gray79, gray80, StillScaryAJP}

In the upper left of Fig.\ \ref{fig:multipoles from dipoles}, we see the simplest magnetic system: a dipole.\cite{griffiths_2017}  Two magnetic dipoles facing each other will cancel the dipole moment in the multipole expansion, leaving the quadrupole moment as the leading term in the expansion (middle and lower left panels).  On the upper right side of Fig.\ \ref{fig:multipoles from dipoles}, the planar quadrupole used in particle accelerators is simply the combination of two linear quadrupoles, again leaving the quadrupole moment as the leading term in the magnetic field.  In the Supplementary Material,\cite{supplementary} we derive the fact that six dipoles arranged as shown for the sextupole configuration will cancel the dipole \textbf{and} quadrupole moments, leaving the octopole moment as the leading term.  Similarly the octupole magnet will cancel the dipole, quadrupole, and octopole terms, thus leading term is the hexadecapole (16-pole) moment.

In order for the magnetic field to be dominated by a single term in the multipole expansion, we must look at distances where
$r$ from the center of the configuration to the point of interest is much greater than the radius $a$ of the polygon, $r \gg a$. If we choose the point of interest in a 2D plane perpendicular to the faces of the magnets and lying along the line that extends from the center of the configuration through one of the faces, then by symmetry we expect the net magnetic field from the magnet configuration to only have a radial field component.  In the Supplementary Material,\cite{supplementary} we derive the magnetic field and show that the magnetic field of the multipole configuration follows a pattern, given by:
\begin{equation}\label{eq:generalfield}
    B = f\, \frac{\mu_0 }{4 \pi}\, \frac{m a^{l-1}}{r^{l + 2}}
\end{equation}
where $m$ is the dipole moment, $f$ is a numerical prefactor that depends on the magnet configuration and the choice of axis,  $l$ is the order of the leading multipole moment, and $a$ is the distance from the center of the configuration to the location of one of the dipoles. The leading multipole moment and the numerical values of the prefactor $f$ are given in Table \ref{tab:prefactors}.  

\begin{table}\caption{Outside a multipole magnet configuration, the power in $B \propto 1/r^{l+2}$ is determined by the multipole order $l$.  The prefactor determines the value of the magnetic field generated by multipole configurations given by Eq.\ \ref{eq:generalfield}.  The magnetic field is calculated at a distance $r$ from the center of the configuration along a line from the center through one of the dipoles, except for the linear quadrupole perpendicular to the axis, which is along a line perpendicular to that line.\cite{octupole}}
\begin{tabular}{l c c c  }
\begin{tabular}[c]{@{}l@{}}Multipole\\ Magnet \end{tabular}  & \begin{tabular}[c]{@{}c@{}}Leading\\ Mutipole Moment\end{tabular} &  \begin{tabular}[c]{@{}c@{}}Multpole\\ Order $l$\end{tabular}   & \begin{tabular}[c]{@{}c@{}}Prefactor\\ $f$\end{tabular}  \\ \hline
Dipole                                                                             & Dipole     & 1 & 2               \\ 
\begin{tabular}[c]{@{}l@{}}Linear Quadrupole,\\ perpendicular to axis\end{tabular} & Quadrupole & 2 & 6               \\ 
\begin{tabular}[c]{@{}l@{}}Linear Quadrupole,\\ parallel to axis\end{tabular}      & Quadrupole & 2 & 12              \\ 
Planar Quadrupole                                                                  & Quadrupole & 2 & 18              \\ 
Sextupole                                                                          & Octopole   & 3 & 45              \\ 
Octupole                                                                           & Hexadecapole & 4 & 87.5 \\ \hline
\end{tabular}\label{tab:prefactors}
\end{table}

Eq.\ \ref{eq:generalfield} is a simple formula that conforms to all expectations of how the field exterior to the multipole magnets should behave:\ the power $l+2$ increases as the leading multipole moment increases, additional factors with units of length appear in the numerator to ensure the dimensions are correct, the field depends on the strength of the dipoles used to make the multipole magnet and only the prefactor in front changes. Armed with this simplified model and expression for the magnetic field, the curious student is well-positioned to delve into a visual and experimental exploration of the fields generated by the accelerator magnet configurations.

\section{Visualization via Simulation}
For many students, curiosity first demands that they see the phenomena they are studying in action.  
We use Python and the open-source library Magylib,\cite{magpylib2020}  a package that can model macroscopic permanent magnets of various sizes and strengths, position them in space, evolve them in time, and calculate the resultant magnetic field in space and time. This package is free and user-friendly,  and it offers students excellent introductory-level practice with programming, magnets, and magnetic fields.

\begin{figure}
   \centering
   \includegraphics[width=0.8\linewidth]{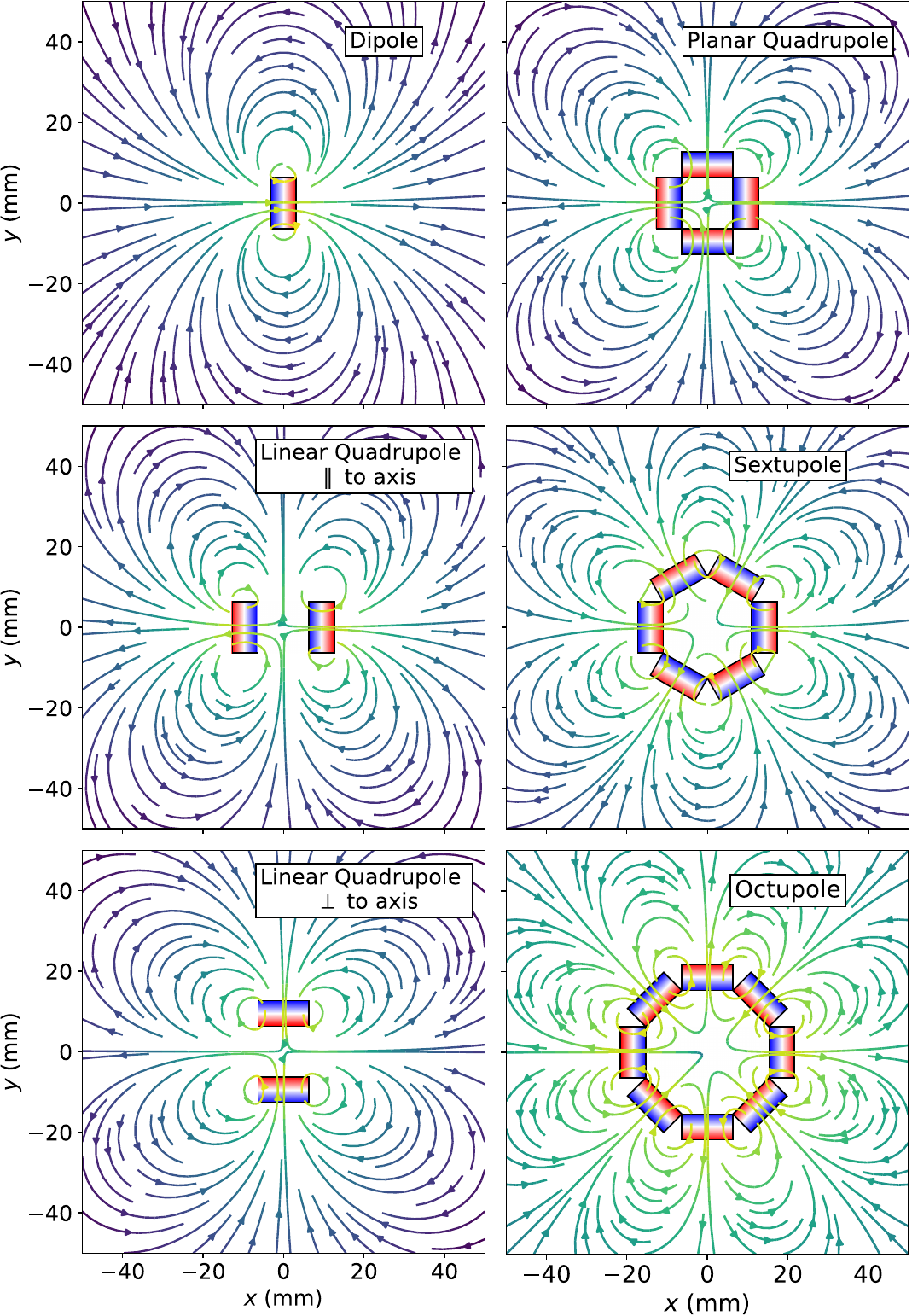}
   \caption{Field lines for all five magnet configurations. Permanent magnets are drawn to scale, with north and south poles shaded red and blue, respectively. The field lines are drawn with arrows, and the color of the field line indicates the magnitude of the magnetic field, varying in strength from white to yellow to red.  Multipole magnet naming convention follows the standard in accelerator physics.\cite{octupole}}
   \label{fig: field contours}
\end{figure}

Students can use the Magpylib package to visualize the magnetic field generated by the permanent magnet configurations shown in Fig.\ \ref{fig:multipoles from dipoles}. In Fig.\ \ref{fig: field contours}, each of the permanent magnets is drawn to scale with north (red) and south (blue) poles indicated on the diagram.  The size $a$ of the configuration varies from $a=9.5$ mm (quadrupoles) to $a=14.2$ mm (sextupole) to $a=18.5$ mm (octupole).\cite{octupole} The simulation field lines agree well with the known fields for dipoles\cite{griffiths_2017} and quadrupoles.\cite{gray09}  Our Magpylib simulation code is available online and in the Supplementary Material.\cite{EPAPS, supplementary}

Note the symmetry in the field patterns. The dipole and the linear quadrupole magnets create familiar two- and four-lobed field patterns, and the number of lobes equals the total number of magnetic poles present.  The orientations of the accelerator magnet analogs (right side of Fig.\ \ref{fig: field contours}), meanwhile, create symmetries where the number of lobes in each field pattern is the same as the number of \textit{magnets} present or the number of poles that face inward towards the center.  While this differs from the more common explanations for a dipole and quadrupole, it can be explained through a careful examination of the visualizations.  As the field leaves each north (red) pole, the field will bend towards the nearest south (blue) pole, which are directly adjacent to the north pole magnet.  This repeats with every north pole, giving two lobes per outward facing north pole.  

With this visualization students can link the field symmetry to the expected dominant term in the multipole expansion (Table \ref{tab:prefactors}).  In addition, in all patterns, one can see that the field at a point on the line from the center of the magnet configuration through the center of any magnet points along that line (i.e., in the $r$-direction only), as predicted in Sec.\ \ref{sec:analytical}.  Finally, students can use the visualization of the dipole magnet to guide their data collection and analysis of the location of the sensor within the phone, as discussed in the Supplementary Material.\cite{supplementary}

\section{Experiment Design}
The experiment presented in this article includes many elements available to engage students in experiment design, fabrication, and setup.  The complete apparatus is inexpensive and the setup and data-taking can be completed in-person or remotely.

For our experiments, we purchased NdFeB N52 cuboid magnets\cite{CMSmags} of size $12.7 \times 12.7 \times 6.35$ \si{mm^3}.  We designed magnet holders (available online\cite{EPAPS,supplementary})  for each magnet configuration and 3D printed the holders at Shapeways.\cite{Shapeways}  These magnet holders are necessary to ensure the magnets are in the proper position to eliminate lower-order multipole moments.  

We also designed a smartphone bed that can be adjusted to accommodate all smartphones currently available, and an extension arm to position the magnet holders and prevent rotation.\cite{EPAPS, supplementary}  The extension arm has side walls that can be added or removed to accommodate the holders for the different multipole magnet configurations and can slide along the smartphone bed to align the arm to the magnetic field sensor inside the phone. Fig.\ \ref{fig:overview} shows the physical apparatus (smartphone bed and extension arm) and the sextupole magnet holder. The bed and extension arm are not required, but they simplify the data collection in two ways:\ first, they accurately locate the magnetic field sensor inside the phone, and second, they facilitate a reliable and repeatable measurement of the distance from the center of the multipole configuration to the sensor.  
Without the bed and extension arm, students are free to imagine how to conduct the measurement.  Our remote students in fall 2020 had many different and clever experimental setups, most made from cardboard and wood.

Lastly, this experiment requires a smartphone and an app to be able to access the magnetic field measurements made by the smartphone.  There are several cross-platform apps available to measure the magnetic field.  We used Phyphox,\cite{Phyphox} which works on Android and iOS devices, is free, can export the data as a .csv file, offers both calibrated and raw magnetometry data, and can be operated remotely from a computer via WiFi connection.  The total cost of the experimental setup, excluding the smartphone, 
was less than \$250.

Here we summarize the experiment; full details for each step are discussed in the Supplementary Material.\cite{supplementary} To collect data, we first located the magnetic field sensor inside the smartphone.  Once located, we measured the magnetic field as a function of distance for the different magnet configurations.  Each magnet configuration was placed at fixed locations along the extension arm, and the distance from the center of the configuration to the sensor was measured in CAD. For all measurements, the residual magnetic field measured without magnets present was subtracted from the measurement.

\section{Data and Analysis}
\begin{figure}
   \centering
   \includegraphics[width=0.95\linewidth]{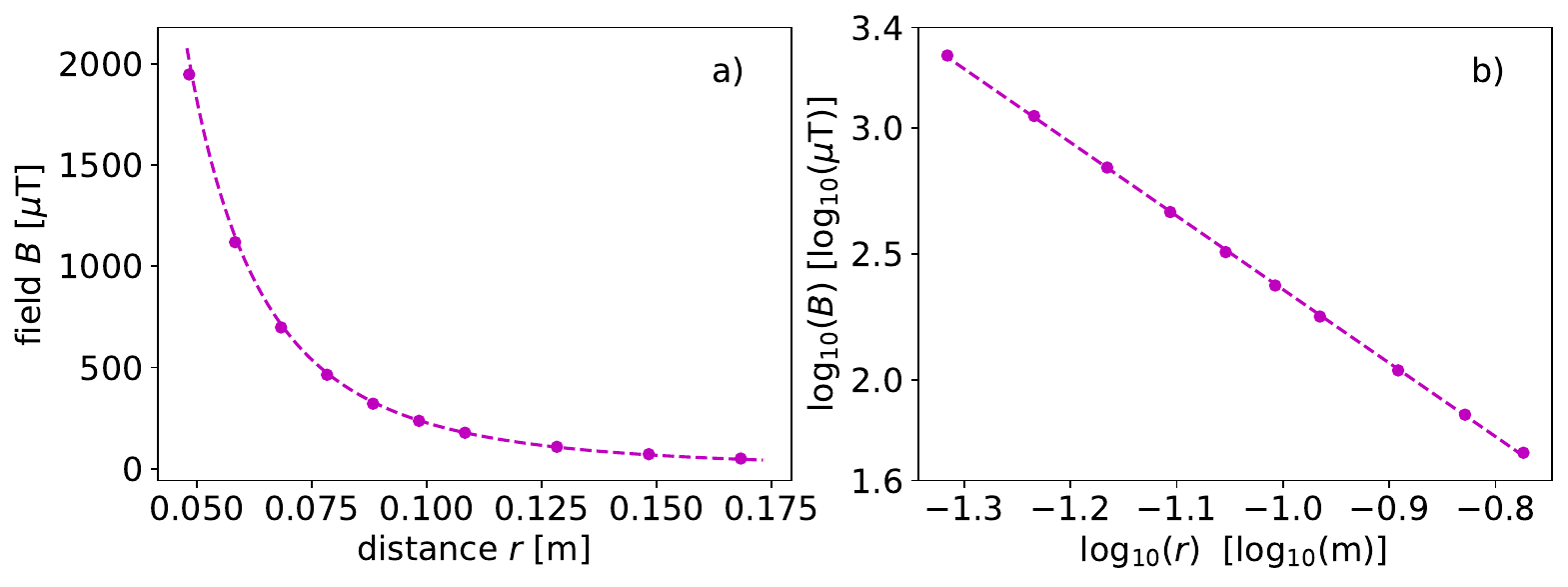}
   \caption{Measurements of the magnetic field generated by a single permanent magnet. (a) shows the magnetic field as a function of distance with an inverse cubic fit.  (b) is a log-log plot showing clear power-law behavior.  The fit line has a slope of $-2.97 \pm 0.01$.  The intercept is used to determine the magnitude of the dipole moment $m$.}
   \label{fig:dipole}
\end{figure}

The results for a single magnet (dipole) are shown as dots in Fig.\ \ref{fig:dipole} (uncertainty in the data is much smaller than the points).   In Fig.\ \ref{fig:dipole}(a), the dotted lines represent a fitted theoretical curve, found by fixing the slope at $-3$ and allowing the least-squares fit to vary the dipole moment to find an appropriate fit to the model.   Taking the logarithm of both sides of Eq.\ \ref{eq:generalfield} gives us:
\begin{equation}\label{eq:loggeneralfield}
    \log(B) = \log\left(\frac{\mu_0}{4 \pi} f m a^{l-1}\right) - (l+2)\log(r),
\end{equation}
where the power becomes the slope of a log-log plot.  Fig.\ \ref{fig:dipole}(b) shows a log-log plot of the same data with a linear least-squares fit where the power is allowed to vary, yielding a slope of $-2.97 \pm 0.01$.\cite{loglog} This slope corresponds to $l = 1$ and confirms that the dipole moment dominates the behavior of the magnetic field.  The dipole moment $m$ in Eq.\ \ref{eq:generalfield} can be found using the intercept in Fig.\ \ref{fig:dipole}(b) (using the value for $f$ from Table \ref{tab:prefactors}).  The intercept of $-0.61 \pm 0.01$ gives a dipole moment of $m = 1.21 \pm 0.02$ \SI{}{\ampere \meter^2}. 

\begin{figure}
   \centering
   \includegraphics[width=0.95\linewidth]{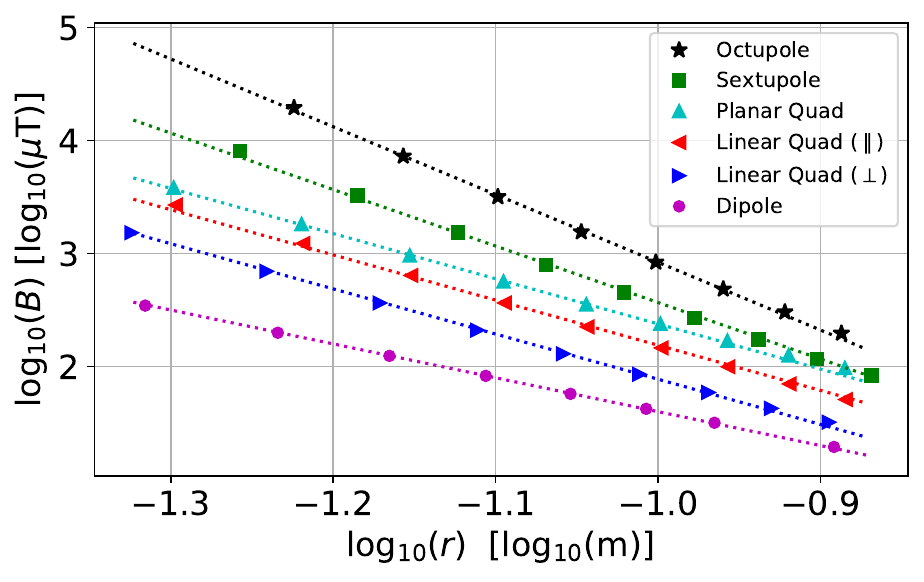}
   \caption{Measurements on all multipole magnet configurations.  The dotted lines are fits to the model from Eq.\ \ref{eq:generalfield} using Table \ref{tab:prefactors}.  The only free parameter in the model line fits is the dipole moment $m$, which was found via least squares fitting, and varies between 1.21 and 1.51 \SI{}{\ampere \meter^2}.  These results show that the experimental data from the multipole magnet configurations agree well with the model for both slope and intercept, thus the magnet configurations are well modeled by ideal dipoles and the magnetic field multipole field expansion.  For clarity, the data for the dipole is offset vertically by $-0.75$, the sextupole by $+0.5$, and the octupole by $+1$. }
   \label{fig:all}
\end{figure}

For the linear quadrupole oriented in two ways and the planar quadrupole, we expect the field to vary as $B \propto 1/r^4$, with the different prefactors as listed in Table \ref{tab:prefactors}.  The experimental results are shown as markers in Fig.\ \ref{fig:all} (uncertainty in the data is much smaller than the marks).  The dotted lines represent theoretical curves, found by fixing the slope at $-4$. The fits are quite good by eye, and the dipole moments are \SI{1.35 \pm 0.07}{\ampere \m^2} for the linear quadrupole parallel to the axis, \SI{1.35 \pm 0.06}{\ampere \m^2} for the linear quadrupole perpendicular to the axis, and \SI{1.39 \pm 0.03}{\ampere \m^2} for the planar quadrupole. The values for $m$ do not vary significantly among the quadrupole magnet configurations, though they are significantly discrepant from the value found for the single dipole magnet.  


For the sextupole and octupole magnet configurations, we again fix the slopes at $-5$ and $-6$, respectively, and allow the dipole moment $m$ to vary in the least-squares fit. The strong correlation between the data and theory verifies the power-law dependence of the fields.  The value for the dipole moment $m$ is determined to be \SI{1.29 \pm 0.08}{\ampere \m^2} and \SI{1.51 \pm 0.08}{\ampere \m^2} for the sextupole and octupole configurations, respectively.  We note that these results for $m$ are again significantly discrepant from the value found for the dipole, yet the largest discrepancy, found for the most complex configuration, the octupole magnet, varies by only 25\%. This is impressive given the need for the dipole, quadrupole, \textit{and} octopole moments in the multipole expansion to cancel out in order to produce a dominant hexadecapole moment that presents as a $1/r^6$ dependency in the magnetic field data. The dipole moments of the magnets are assumed to be equal in the model but vary by $0.5\%$ in the experiment. We also expect that slight misalignments of the magnets in the holder or rotation of the magnet holder will affect the value for the dipole moment.

Our data in Fig.\ \ref{fig:all} represent the full range of useful data collection for the quadruple, sextupole, and octupole magnets. At distances closer than $\sim$ \SI{5}{cm}, the strong permanent magnets can harm the phone (this distance may be smaller for larger or more powerful magnets; students should be careful not to damage their smartphone).   For the dipole magnet, measured data remains consistent with $B \propto 1/r^3$ for distances up to \SI{50}{cm} from the sensor.  By contrast, at distances farther than $\sim$ \SI{15}{cm} the data for the quadrupole, sextupole, and octupole magnet configurations curve up and away from the expected behavior when plotted on a log-log plot. Small differences in the dipole moment $m$ of the permanent magnets means the cancellation isn't perfect and leads to a small net dipole moment of the system.  At large $r$, this remnant dipole moment ($\propto 1/r^3$) dominates over the quadrupole, octopole, or hexadecapole moments, which drop off more rapidly as $\sim 1/r^4$, $1/r^5$ and $1/r^6$, respectively.



\section{Calculations via Simulation}
\begin{figure}
   \centering
   \includegraphics[width=0.85\linewidth]{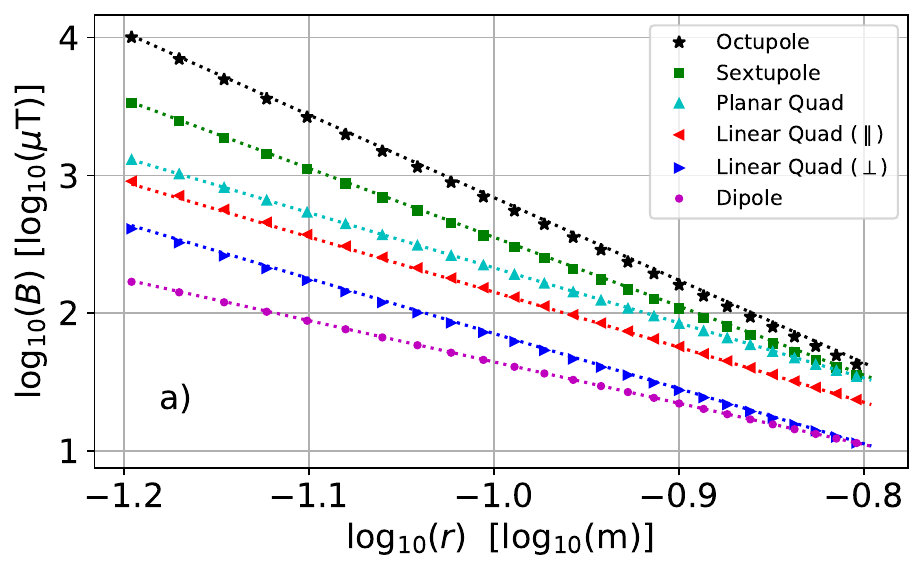}
   \includegraphics[width=0.85\linewidth]{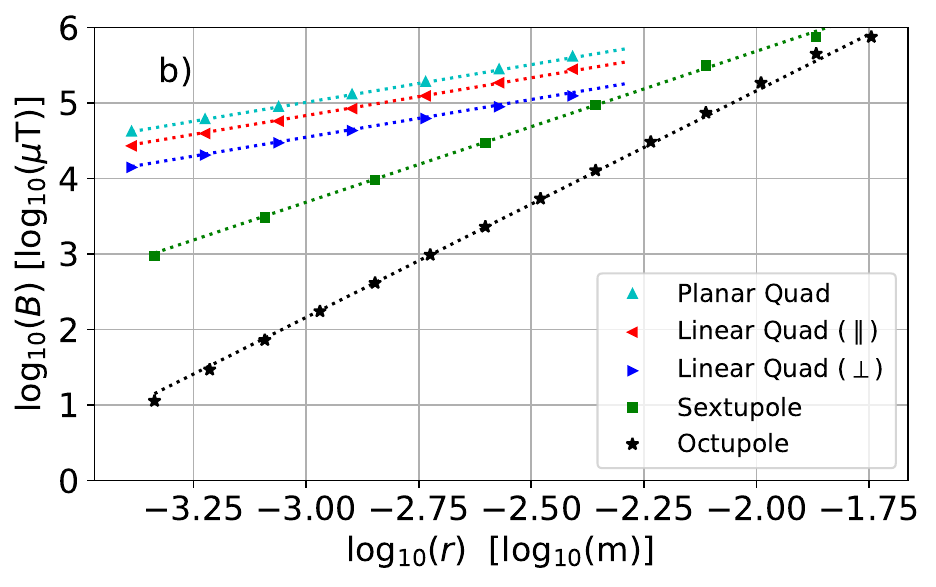}
   \caption{Dots indicate the simulated magnetic field from the magnet configurations shown in Fig.\ \ref{fig: field contours}. In (a), the dotted lines are fits to the model from Eq.\ \ref{eq:generalfield} using Table \ref{tab:prefactors}.  All magnets have the same dipole moment, $m= \SI{1.12}{\ampere \meter^2}$.  We can see the simulation agrees well with the model, showing that our non-interacting permanent magnets behave very similarly to ideal dipoles. The dipole data is offset vertically by $-0.75$; the sextupole by $+0.5$ and the octupole by $+1$.  In (b), we show the contact field inside the magnet configurations.  The fit lines are slopes of $+1$ (quadrupoles), $+2$ (sextupole), and $+3$ (octupole).  The octupole is offset vertically by $-1$. }
   \label{fig:all_sim}
\end{figure}

In addition to visualization of the magnetic field, the Magpylib package can also be used to \textbf{calculate} the magnetic field at any point in space.  If we use the simulation to determine the magnetic field along the positive $x$-axis in each of the configurations in Fig.\ \ref{fig: field contours}, we can reproduce our experimental setup \textit{in silico}.  We can then compare our simulated results to the expected results from Eq.\ \ref{eq:generalfield} using Table \ref{tab:prefactors}.  Our simulated results are compared to the theoretical predictions in Fig.\ \ref{fig:all_sim}(a), where we see excellent agreement between simulation and theory.  All the magnets in the simulation and theoretical fits in Fig.\ \ref{fig:all_sim} have the same dipole moment of $m = \SI{1.12}{\ampere \meter^2}$ (in contrast with our experimental results, where the dipole moment $m$ varied by up to 25\%).  The results in Fig.\ \ref{fig:all_sim} corroborate our analytical predictions and indicate that non-interacting permanent magnets do indeed create the expected multipole fields.

We can extract still more information from the simulation.  We verified the behavior far from the center of the multipole magnet configuration, but we can also examine the magnetic field in regions we cannot access experimentally.  We can examine the field closer than $\sim$ \SI{5}{cm}, where the assumption that $r \gg a$ begins to break down.  This simulation also allows the user to look \textit{inside} the magnet configurations, shown in Fig.\ \ref{fig:all_sim}(b).  In this figure, we can see linear behavior of the field inside the quadrupole configurations and  quadratic behavior inside the sextupole configuration.  This linear and quadratic behavior (slopes of $+1$ and $+2$) is utilized for bending and focusing the particle beam in accelerator physics.\cite{niceaccelerator, CERN_JAIcourse}  The octupole, with cubic behavior (slope of $+3$), is not commonly used in accelerators.

Finally, the simulation can be used to create arbitrary collections of permanent magnets and arrange them in space.  In our application, this means we can create a decapole or dodecapole magnet configuration and extend the simulation to higher multipole moments.  This is very useful, given that the algebra required to calculate the prefactor for the octupole magnet, while not conceptually complex, is certainly complicated and error-prone.

\section{Conclusions}

Using particle accelerators as a ``hook," we draw students into a multifaceted project that provides a broad range of skill-building opportunities through an examination of the magnetic field behavior exterior to a collection of small-scale accelerator magnet analogs.
We have examined five different magnet configurations via theory using ideal magnetic dipoles, via experiment using commercially available NdFeB permanent magnets, and via simulation using Python and Magpylib to simulate permanent magnets.  In all cases, we see that $B \propto 1/r^{l+2}$.  As the multipole magnet increases in complexity from dipole to quadrupole to sextupole to octupole, theory predicts the power $l+2$ should also increase; this is also borne out by our theoretical derivations, experiments, and simulations.  The predicted powers of $l+2 = 3, 4, 5, 6$ for the magnet configurations are also verified in our calculations, simulations, and experimental results. In our experimental fits, the measured values of the dipole moment $m$ varies by 25\% or less.  For the most complex configuration, the octupole magnet, this discrepancy reaches the maximum of 25\%, a remarkably small difference considering the stringent requirements necessary to cancel the dipole, quadrupole, and octopole moments exactly to allow the $B\sim 1/r^6$ hexadecapole moment to dominate the far-field behavior.

The equipment used in this work is highly accessible, given the ubiquity of smartphones and personal computers and the proliferation of user-friendly and free software packages like Python and Magpylib.  
Most students will need to be supplied with only magnets and magnet holders to complete this project. This makes the project ideal for in-person or remote work, making magnetic labs readily available to students far from a physical campus.

This work can be customized to fit a student's needs or interests, be they analytical theory, CAD design, 3D printing, computer visualization and simulation, or data collection and analysis.  The project can be expanded or compressed depending on the course, with different parts of the project appropriate for sophomore-level courses all the way to a senior project.  In addition, this experiment is ripe for additional explorations.  A student interested in experimentation could measure the field as a function of angle, and verify each individual multipole magnet's unique field configuration.  A student motivated by computation could model a non-symmetric arrangement of magnets or investigate the effect of including magnets of differing field strength, either (or both) of which might verify the discrepancy in the measured value of the magnet dipole moment $m$.  Recognizing that permanent magnets are not perfect dipoles, a student engaged by analysis could modify the fit to include both higher and lower-order terms in the expansion, and use residual plots to confirm the necessity of these terms.  A student interested in 3D printing can print the smartphone bed and extension arm, either with or without modifications for their own smartphone.  A student excited by engineering design could change the phone holder such that any location for the extension arm is possible (only discrete locations are available in our experiment).  Finally, a student excited to explore accelerator physics could design and build a larger-scale version, such that the smartphone could fit \textbf{inside} the configuration, and measure the contact fields relevant for accelerators.  \footnote{We are indebted to our editors and anonymous reviewers for many of these suggestions.}

Lastly, the intentional inclusion of low-cost solutions, in our case, the use of smartphone sensors and Python, can be viewed as a beacon of hope.  No longer is the learning of physical phenomena and the scientific method bound to classrooms but is now a universal opportunity.  In times of widespread educational inequities, making experiments and simulations low-cost and accessible to all provides an opportunity to contribute to educational models that can reach underserved communities.

\begin{acknowledgments}
The authors would like to thank the brave remote students of Intermediate Lab during fall 2020 for attempting the first versions of this experiment, and the in-person students in fall 2021 who discovered other needed improvements.  A special acknowledgment goes to Alex Powell, whose determined data collection uncovered errors in our assumptions, and to our colleague Jerome Fung for his careful reading and editing of our manuscript.

The authors would also like to thank the editors and the anonymous reviewers of the manuscript for their careful edits and corrections, which have greatly improved this paper.

The authors have no conflicts of interest to disclose.
\end{acknowledgments}



\begin{thebibliography}{36}%
\makeatletter
\providecommand \@ifxundefined [1]{%
 \@ifx{#1\undefined}
}%
\providecommand \@ifnum [1]{%
 \ifnum #1\expandafter \@firstoftwo
 \else \expandafter \@secondoftwo
 \fi
}%
\providecommand \@ifx [1]{%
 \ifx #1\expandafter \@firstoftwo
 \else \expandafter \@secondoftwo
 \fi
}%
\providecommand \natexlab [1]{#1}%
\providecommand \enquote  [1]{``#1''}%
\providecommand \bibnamefont  [1]{#1}%
\providecommand \bibfnamefont [1]{#1}%
\providecommand \citenamefont [1]{#1}%
\providecommand \href@noop [0]{\@secondoftwo}%
\providecommand \href [0]{\begingroup \@sanitize@url \@href}%
\providecommand \@href[1]{\@@startlink{#1}\@@href}%
\providecommand \@@href[1]{\endgroup#1\@@endlink}%
\providecommand \@sanitize@url [0]{\catcode `\\12\catcode `\$12\catcode
  `\&12\catcode `\#12\catcode `\^12\catcode `\_12\catcode `\%12\relax}%
\providecommand \@@startlink[1]{}%
\providecommand \@@endlink[0]{}%
\providecommand \url  [0]{\begingroup\@sanitize@url \@url }%
\providecommand \@url [1]{\endgroup\@href {#1}{\urlprefix }}%
\providecommand \urlprefix  [0]{URL }%
\providecommand \Eprint [0]{\href }%
\providecommand \doibase [0]{https://doi.org/}%
\providecommand \selectlanguage [0]{\@gobble}%
\providecommand \bibinfo  [0]{\@secondoftwo}%
\providecommand \bibfield  [0]{\@secondoftwo}%
\providecommand \translation [1]{[#1]}%
\providecommand \BibitemOpen [0]{}%
\providecommand \bibitemStop [0]{}%
\providecommand \bibitemNoStop [0]{.\EOS\space}%
\providecommand \EOS [0]{\spacefactor3000\relax}%
\providecommand \BibitemShut  [1]{\csname bibitem#1\endcsname}%
\let\auto@bib@innerbib\@empty
\bibitem [{\citenamefont {Appleby}\ \emph {et~al.}(2020)\citenamefont
  {Appleby}, \citenamefont {Burt}, \citenamefont {Clarke},\ and\ \citenamefont
  {Owen}}]{niceaccelerator}%
  \BibitemOpen
  \bibfield  {author} {\bibinfo {author} {\bibfnamefont {R.}~\bibnamefont
  {Appleby}}, \bibinfo {author} {\bibfnamefont {G.}~\bibnamefont {Burt}},
  \bibinfo {author} {\bibfnamefont {J.}~\bibnamefont {Clarke}},\ and\ \bibinfo
  {author} {\bibfnamefont {H.}~\bibnamefont {Owen}},\ }\enquote {\bibinfo
  {title} {The science and technology of particle accelerators},}\ in\ \href
  {https://doi.org/10.1201/9781351007962} {\emph {\bibinfo {booktitle} {The
  Science and Technology of Particle Accelerators}}}\ (\bibinfo  {publisher}
  {CRC Press},\ \bibinfo {year} {2020})\ pp.\ \bibinfo {pages} {112--162},\
  \bibinfo {edition} {1st}\ ed.\BibitemShut {Stop}%
\bibitem [{\citenamefont {Milanese}(2022)}]{CERN_JAIcourse}%
  \BibitemOpen
  \bibfield  {author} {\bibinfo {author} {\bibfnamefont {A.}~\bibnamefont
  {Milanese}},\ }\href@noop {} {\enquote {\bibinfo {title} {An introduction to
  magnets for accelerators},}\ }\bibinfo {howpublished} {\href{John Adams
  Institute for Accelerator Science - Graduate Accelerator Physics
  Course}{https://indico.cern.ch/event/1101643/contributions/}} (\bibinfo
  {year} {2022}),\ \bibinfo {note} {accessed: 2022-04-06}\BibitemShut {NoStop}%
\bibitem [{wik(2022{\natexlab{a}})}]{wikipedia_dipole}%
  \BibitemOpen
  \href {https://en.wikipedia.org/wiki/Dipole_magnet} {\enquote {\bibinfo
  {title} {Dipole magnet},}\ }\bibinfo {howpublished}
  {\url{https://en.wikipedia.org/wiki/Dipole_magnet}} (\bibinfo {year}
  {2022}{\natexlab{a}}),\ \bibinfo {note} {accessed: 2022-05-13}\BibitemShut
  {NoStop}%
\bibitem [{wik(2022{\natexlab{b}})}]{wikipedia_quad}%
  \BibitemOpen
  \href {https://en.wikipedia.org/wiki/Quadrupole_magnet} {\enquote {\bibinfo
  {title} {Quadrupole magnet},}\ }\bibinfo {howpublished}
  {\url{https://en.wikipedia.org/wiki/Quadrupole_magnet}} (\bibinfo {year}
  {2022}{\natexlab{b}}),\ \bibinfo {note} {accessed: 2022-05-13}\BibitemShut
  {NoStop}%
\bibitem [{wik(2022{\natexlab{c}})}]{wikipedia_sext}%
  \BibitemOpen
  \href {https://en.wikipedia.org/wiki/Sextupole_magnet} {\enquote {\bibinfo
  {title} {Sextupole magnet},}\ }\bibinfo {howpublished}
  {\url{https://en.wikipedia.org/wiki/Sextupole_magnet}} (\bibinfo {year}
  {2022}{\natexlab{c}}),\ \bibinfo {note} {accessed: 2022-05-13}\BibitemShut
  {NoStop}%
\bibitem [{\citenamefont {Gray}, \citenamefont {Karl},\ and\ \citenamefont
  {Novikov}(2009)}]{gray09}%
  \BibitemOpen
  \bibfield  {author} {\bibinfo {author} {\bibfnamefont {C.~G.}\ \bibnamefont
  {Gray}}, \bibinfo {author} {\bibfnamefont {G.}~\bibnamefont {Karl}},\ and\
  \bibinfo {author} {\bibfnamefont {V.~A.}\ \bibnamefont {Novikov}},\
  }\bibfield  {title} {\enquote {\bibinfo {title} {Quadrupolar contact fields:
  Theory and applications},}\ }\href {https://doi.org/10.1119/1.3138700}
  {\bibfield  {journal} {\bibinfo  {journal} {American Journal of Physics}\
  }\textbf {\bibinfo {volume} {77}},\ \bibinfo {pages} {807--817} (\bibinfo
  {year} {2009})},\ \Eprint
  {https://arxiv.org/abs/https://doi.org/10.1119/1.3138700}
  {https://doi.org/10.1119/1.3138700} \BibitemShut {NoStop}%
\bibitem [{\citenamefont {Gray}, \citenamefont {Karl},\ and\ \citenamefont
  {Novikov}(2010)}]{gray10}%
  \BibitemOpen
  \bibfield  {author} {\bibinfo {author} {\bibfnamefont {C.~G.}\ \bibnamefont
  {Gray}}, \bibinfo {author} {\bibfnamefont {G.}~\bibnamefont {Karl}},\ and\
  \bibinfo {author} {\bibfnamefont {V.~A.}\ \bibnamefont {Novikov}},\
  }\bibfield  {title} {\enquote {\bibinfo {title} {Magnetic multipolar contact
  fields: The anapole and related moments},}\ }\href
  {https://doi.org/10.1119/1.3427412} {\bibfield  {journal} {\bibinfo
  {journal} {American Journal of Physics}\ }\textbf {\bibinfo {volume} {78}},\
  \bibinfo {pages} {936--948} (\bibinfo {year} {2010})},\ \Eprint
  {https://arxiv.org/abs/https://doi.org/10.1119/1.3427412}
  {https://doi.org/10.1119/1.3427412} \BibitemShut {NoStop}%
\bibitem [{\citenamefont {Monteiro}\ and\ \citenamefont
  {MartÃ­}(2022)}]{smartphone_resource}%
  \BibitemOpen
  \bibfield  {author} {\bibinfo {author} {\bibfnamefont {M.}~\bibnamefont
  {Monteiro}}\ and\ \bibinfo {author} {\bibfnamefont {A.~C.}\ \bibnamefont
  {MartÃ­}},\ }\bibfield  {title} {\enquote {\bibinfo {title} {Resource letter
  {MDS}-1: {M}obile devices and sensors for physics teaching},}\ }\href
  {https://doi.org/10.1119/5.0073317} {\bibfield  {journal} {\bibinfo
  {journal} {American Journal of Physics}\ }\textbf {\bibinfo {volume} {90}},\
  \bibinfo {pages} {328--343} (\bibinfo {year} {2022})},\ \Eprint
  {https://arxiv.org/abs/https://doi.org/10.1119/5.0073317}
  {https://doi.org/10.1119/5.0073317} \BibitemShut {NoStop}%
\bibitem [{\citenamefont {O'Brien}(2021)}]{elearningAJP}%
  \BibitemOpen
  \bibfield  {author} {\bibinfo {author} {\bibfnamefont {D.~J.}\ \bibnamefont
  {O'Brien}},\ }\bibfield  {title} {\enquote {\bibinfo {title} {A guide for
  incorporating e-teaching of physics in a post-covid world},}\ }\href
  {https://doi.org/10.1119/10.0002437} {\bibfield  {journal} {\bibinfo
  {journal} {American Journal of Physics}\ }\textbf {\bibinfo {volume} {89}},\
  \bibinfo {pages} {403--412} (\bibinfo {year} {2021})},\ \Eprint
  {https://arxiv.org/abs/https://doi.org/10.1119/10.0002437}
  {https://doi.org/10.1119/10.0002437} \BibitemShut {NoStop}%
\bibitem [{man()}]{manypapers}%
  \BibitemOpen
  \href@noop {} {}\bibinfo {note} {See Ref.\ \onlinecite{smartphone_resource}
  (and references therein) for additional references.}\BibitemShut {Stop}%
\bibitem [{\citenamefont {Pili}\ and\ \citenamefont
  {Violanda}(2018)}]{RotationswithBTPT}%
  \BibitemOpen
  \bibfield  {author} {\bibinfo {author} {\bibfnamefont {U.}~\bibnamefont
  {Pili}}\ and\ \bibinfo {author} {\bibfnamefont {R.}~\bibnamefont
  {Violanda}},\ }\bibfield  {title} {\enquote {\bibinfo {title} {Measuring
  average angular velocity with a smartphone magnetic field sensor},}\ }\href
  {https://doi.org/10.1119/1.5021442} {\bibfield  {journal} {\bibinfo
  {journal} {The Physics Teacher}\ }\textbf {\bibinfo {volume} {56}},\ \bibinfo
  {pages} {114--115} (\bibinfo {year} {2018})},\ \Eprint
  {https://arxiv.org/abs/https://doi.org/10.1119/1.5021442}
  {https://doi.org/10.1119/1.5021442} \BibitemShut {NoStop}%
\bibitem [{\citenamefont {Arabasi}\ and\ \citenamefont
  {Al-Taani}(2016)}]{MeasEarthBEJP}%
  \BibitemOpen
  \bibfield  {author} {\bibinfo {author} {\bibfnamefont {S.}~\bibnamefont
  {Arabasi}}\ and\ \bibinfo {author} {\bibfnamefont {H.}~\bibnamefont
  {Al-Taani}},\ }\bibfield  {title} {\enquote {\bibinfo {title} {Measuring the
  earth's magnetic field dip angle using a smartphone-aided setup: a simple
  experiment for introductory physics laboratories},}\ }\href
  {https://doi.org/10.1088/1361-6404/38/2/025201} {\bibfield  {journal}
  {\bibinfo  {journal} {European Journal of Physics}\ }\textbf {\bibinfo
  {volume} {38}},\ \bibinfo {pages} {025201} (\bibinfo {year}
  {2016})}\BibitemShut {NoStop}%
\bibitem [{\citenamefont {Tronicke}\ and\ \citenamefont
  {Trauth}(2018)}]{BfieldsinclassEJP}%
  \BibitemOpen
  \bibfield  {author} {\bibinfo {author} {\bibfnamefont {J.}~\bibnamefont
  {Tronicke}}\ and\ \bibinfo {author} {\bibfnamefont {M.~H.}\ \bibnamefont
  {Trauth}},\ }\bibfield  {title} {\enquote {\bibinfo {title} {Classroom-sized
  geophysical experiments: magnetic surveying using modern smartphone
  devices},}\ }\href {https://doi.org/10.1088/1361-6404/aaad5b} {\bibfield
  {journal} {\bibinfo  {journal} {European Journal of Physics}\ }\textbf
  {\bibinfo {volume} {39}},\ \bibinfo {pages} {035806} (\bibinfo {year}
  {2018})}\BibitemShut {NoStop}%
\bibitem [{\citenamefont {Ogawara}, \citenamefont {Bhari},\ and\ \citenamefont
  {Mahrley}(2017)}]{simplemeasTPT}%
  \BibitemOpen
  \bibfield  {author} {\bibinfo {author} {\bibfnamefont {Y.}~\bibnamefont
  {Ogawara}}, \bibinfo {author} {\bibfnamefont {S.}~\bibnamefont {Bhari}},\
  and\ \bibinfo {author} {\bibfnamefont {S.}~\bibnamefont {Mahrley}},\
  }\bibfield  {title} {\enquote {\bibinfo {title} {Observation of the magnetic
  field using a smartphone},}\ }\href {https://doi.org/10.1119/1.4976667}
  {\bibfield  {journal} {\bibinfo  {journal} {The Physics Teacher}\ }\textbf
  {\bibinfo {volume} {55}},\ \bibinfo {pages} {184--185} (\bibinfo {year}
  {2017})},\ \Eprint {https://arxiv.org/abs/https://doi.org/10.1119/1.4976667}
  {https://doi.org/10.1119/1.4976667} \BibitemShut {NoStop}%
\bibitem [{\citenamefont {Shakur}\ and\ \citenamefont
  {Valliant}(2020)}]{HelmholtzMeasTPT}%
  \BibitemOpen
  \bibfield  {author} {\bibinfo {author} {\bibfnamefont {A.}~\bibnamefont
  {Shakur}}\ and\ \bibinfo {author} {\bibfnamefont {B.}~\bibnamefont
  {Valliant}},\ }\bibfield  {title} {\enquote {\bibinfo {title} {Flyby
  measurement of the magnetic field of a helmholtz coil with a smart cart},}\
  }\href {https://doi.org/10.1119/1.5141963} {\bibfield  {journal} {\bibinfo
  {journal} {The Physics Teacher}\ }\textbf {\bibinfo {volume} {58}},\ \bibinfo
  {pages} {15--17} (\bibinfo {year} {2020})},\ \Eprint
  {https://arxiv.org/abs/https://doi.org/10.1119/1.5141963}
  {https://doi.org/10.1119/1.5141963} \BibitemShut {NoStop}%
\bibitem [{\citenamefont {Arribas}\ \emph {et~al.}(2015)\citenamefont
  {Arribas}, \citenamefont {Escobar}, \citenamefont {Suarez}, \citenamefont
  {Najera},\ and\ \citenamefont {Bel{\'{e}}ndez}}]{arribas2015measurement}%
  \BibitemOpen
  \bibfield  {author} {\bibinfo {author} {\bibfnamefont {E.}~\bibnamefont
  {Arribas}}, \bibinfo {author} {\bibfnamefont {I.}~\bibnamefont {Escobar}},
  \bibinfo {author} {\bibfnamefont {C.~P.}\ \bibnamefont {Suarez}}, \bibinfo
  {author} {\bibfnamefont {A.}~\bibnamefont {Najera}},\ and\ \bibinfo {author}
  {\bibfnamefont {A.}~\bibnamefont {Bel{\'{e}}ndez}},\ }\bibfield  {title}
  {\enquote {\bibinfo {title} {Measurement of the magnetic field of small
  magnets with a smartphone: a very economical laboratory practice for
  introductory physics courses},}\ }\href
  {https://doi.org/10.1088/0143-0807/36/6/065002} {\bibfield  {journal}
  {\bibinfo  {journal} {European Journal of Physics}\ }\textbf {\bibinfo
  {volume} {36}},\ \bibinfo {pages} {065002} (\bibinfo {year}
  {2015})}\BibitemShut {NoStop}%
\bibitem [{\citenamefont {Arribas}\ \emph {et~al.}(2020)\citenamefont
  {Arribas}, \citenamefont {Escobar}, \citenamefont {Ramirez-Vazquez},
  \citenamefont {del Pilar Suarez~Rodriguez}, \citenamefont {Gonzalez-Rubio},\
  and\ \citenamefont {Bel{\'{e}}ndez}}]{arribas2020linear}%
  \BibitemOpen
  \bibfield  {author} {\bibinfo {author} {\bibfnamefont {E.}~\bibnamefont
  {Arribas}}, \bibinfo {author} {\bibfnamefont {I.}~\bibnamefont {Escobar}},
  \bibinfo {author} {\bibfnamefont {R.}~\bibnamefont {Ramirez-Vazquez}},
  \bibinfo {author} {\bibfnamefont {C.}~\bibnamefont {del Pilar
  Suarez~Rodriguez}}, \bibinfo {author} {\bibfnamefont {J.}~\bibnamefont
  {Gonzalez-Rubio}},\ and\ \bibinfo {author} {\bibfnamefont {A.}~\bibnamefont
  {Bel{\'{e}}ndez}},\ }\bibfield  {title} {\enquote {\bibinfo {title} {Linear
  quadrupole magnetic field measured with a smartphone},}\ }\href
  {https://doi.org/10.1119/1.5145411} {\bibfield  {journal} {\bibinfo
  {journal} {The Physics Teacher}\ }\textbf {\bibinfo {volume} {58}},\ \bibinfo
  {pages} {182--185} (\bibinfo {year} {2020})},\ \Eprint
  {https://arxiv.org/abs/https://doi.org/10.1119/1.5145411}
  {https://doi.org/10.1119/1.5145411} \BibitemShut {NoStop}%
\bibitem [{\citenamefont {Ortner}\ and\ \citenamefont
  {Coliado~Bandeira}(2020)}]{magpylib2020}%
  \BibitemOpen
  \bibfield  {author} {\bibinfo {author} {\bibfnamefont {M.}~\bibnamefont
  {Ortner}}\ and\ \bibinfo {author} {\bibfnamefont {L.~G.}\ \bibnamefont
  {Coliado~Bandeira}},\ }\bibfield  {title} {\enquote {\bibinfo {title}
  {Magpylib: A free python package for magnetic field computation},}\ }\href
  {https://doi.org/10.1016/j.softx.2020.100466} {\bibfield  {journal} {\bibinfo
   {journal} {SoftwareX}\ } (\bibinfo {year} {2020}),\
  10.1016/j.softx.2020.100466}\BibitemShut {NoStop}%
\bibitem [{\citenamefont {Griffiths}(2017)}]{griffiths_2017}%
  \BibitemOpen
  \bibfield  {author} {\bibinfo {author} {\bibfnamefont {D.~J.}\ \bibnamefont
  {Griffiths}},\ }\enquote {\bibinfo {title} {Electrodynamics},}\ in\ \href
  {https://doi.org/10.1017/9781108333511.009} {\emph {\bibinfo {booktitle}
  {Introduction to Electrodynamics}}}\ (\bibinfo  {publisher} {Cambridge
  University Press},\ \bibinfo {year} {2017})\ pp.\ \bibinfo {pages} {151--158,
  252--266},\ \bibinfo {edition} {4th}\ ed.\BibitemShut {Stop}%
\bibitem [{\citenamefont {McMurry}\ and\ \citenamefont {Fay}(2014)}]{chembook}%
  \BibitemOpen
  \bibfield  {author} {\bibinfo {author} {\bibfnamefont {J.~E.}\ \bibnamefont
  {McMurry}}\ and\ \bibinfo {author} {\bibfnamefont {R.~C.}\ \bibnamefont
  {Fay}},\ }\enquote {\bibinfo {title} {General chemistry: Atoms first},}\ \
  (\bibinfo  {publisher} {Pearson Education},\ \bibinfo {year} {2014})\
  Chap.~\bibinfo {chapter} {10},\ \bibinfo {edition} {2nd}\ ed.\BibitemShut
  {Stop}%
\bibitem [{EPA()}]{EPAPS}%
  \BibitemOpen
  \href@noop {} {}\bibinfo {note} {The CAD drawings and Magpylib simulation
  code are available for download from Github at
  \href{https://github.com/matthew-c-sullivan/MagneticMultipoleExpt}{github.com/matthew-c-sullivan/MagneticMultipoleExpt}.]}\BibitemShut
  {NoStop}%
\bibitem [{sup()}]{supplementary}%
  \BibitemOpen
  \href@noop {} {}\bibinfo {note} {See Supplementary Material at [url to be
  inserted by AIPP].}\BibitemShut {Stop}%
\bibitem [{Note1()}]{Note1}%
  \BibitemOpen
  \bibinfo {note} {A dipole is a current loop of area $A$ and current $I$ with
  dipole moment $m=IA$. An ideal dipole is a loop of infinitesimal area $A$ and
  infinite current $I$ where the product $m=IA$ remains constant.\cite
  {griffiths_2017}}\BibitemShut {NoStop}%
\bibitem [{oct()}]{octupole}%
  \BibitemOpen
  \href@noop {} {}\bibinfo {note} {We are using the spelling that is most
  common in the different fields. ``Octupole'' magnet is the spelling used in
  accelerator physics for the accelerator magnet configuration with eight
  magnets on an octagon. ``Octopole'' moment is the spelling used in
  electrodynamics for the $l=3$ order in the field expansion that varies as
  $1/r^5$.}\BibitemShut {Stop}%
\bibitem [{\citenamefont {Raab}\ and\ \citenamefont
  {de~Lange}(2004)}]{HardMultipoleBook}%
  \BibitemOpen
  \bibfield  {author} {\bibinfo {author} {\bibfnamefont {R.}~\bibnamefont
  {Raab}}\ and\ \bibinfo {author} {\bibfnamefont {O.}~\bibnamefont
  {de~Lange}},\ }\href@noop {} {\emph {\bibinfo {title} {Multipole Theory in
  Electromagnetism: Classical, quantum, and symmetry aspects, with
  applications}}}\ (\bibinfo  {publisher} {Oxford University Press},\ \bibinfo
  {year} {2004})\BibitemShut {NoStop}%
\bibitem [{\citenamefont {Wolski}(2019)}]{wolski2019maxwells}%
  \BibitemOpen
  \bibfield  {author} {\bibinfo {author} {\bibfnamefont {A.}~\bibnamefont
  {Wolski}},\ }\href@noop {} {\enquote {\bibinfo {title} {Maxwell's equations
  for magnets},}\ } (\bibinfo {year} {2019}),\ \Eprint
  {https://arxiv.org/abs/1103.0713} {arXiv:1103.0713 [physics.acc-ph]}
  \BibitemShut {NoStop}%
\bibitem [{\citenamefont {Gray}(1978{\natexlab{a}})}]{gray78}%
  \BibitemOpen
  \bibfield  {author} {\bibinfo {author} {\bibfnamefont {C.~G.}\ \bibnamefont
  {Gray}},\ }\bibfield  {title} {\enquote {\bibinfo {title} {Multipole
  expansions of electromagnetic fields using debye potentials},}\ }\href
  {https://doi.org/10.1119/1.11364} {\bibfield  {journal} {\bibinfo  {journal}
  {American Journal of Physics}\ }\textbf {\bibinfo {volume} {46}},\ \bibinfo
  {pages} {169--179} (\bibinfo {year} {1978}{\natexlab{a}})},\ \Eprint
  {https://arxiv.org/abs/https://doi.org/10.1119/1.11364}
  {https://doi.org/10.1119/1.11364} \BibitemShut {NoStop}%
\bibitem [{\citenamefont {Gray}(1978{\natexlab{b}})}]{HardMultipoleAJP}%
  \BibitemOpen
  \bibfield  {author} {\bibinfo {author} {\bibfnamefont {C.~G.}\ \bibnamefont
  {Gray}},\ }\bibfield  {title} {\enquote {\bibinfo {title} {Simplified
  derivation of the magnetostatic multipole expansion using the scalar
  potential},}\ }\href {https://doi.org/10.1119/1.11285} {\bibfield  {journal}
  {\bibinfo  {journal} {American Journal of Physics}\ }\textbf {\bibinfo
  {volume} {46}},\ \bibinfo {pages} {582--583} (\bibinfo {year}
  {1978}{\natexlab{b}})},\ \Eprint
  {https://arxiv.org/abs/https://doi.org/10.1119/1.11285}
  {https://doi.org/10.1119/1.11285} \BibitemShut {NoStop}%
\bibitem [{\citenamefont {Gray}(1979)}]{gray79}%
  \BibitemOpen
  \bibfield  {author} {\bibinfo {author} {\bibfnamefont {C.~G.}\ \bibnamefont
  {Gray}},\ }\bibfield  {title} {\enquote {\bibinfo {title} {Magnetic multipole
  expansions using the scalar potential},}\ }\href
  {https://doi.org/10.1119/1.11816} {\bibfield  {journal} {\bibinfo  {journal}
  {American Journal of Physics}\ }\textbf {\bibinfo {volume} {47}},\ \bibinfo
  {pages} {457--459} (\bibinfo {year} {1979})},\ \Eprint
  {https://arxiv.org/abs/https://doi.org/10.1119/1.11816}
  {https://doi.org/10.1119/1.11816} \BibitemShut {NoStop}%
\bibitem [{\citenamefont {Gray}(1980)}]{gray80}%
  \BibitemOpen
  \bibfield  {author} {\bibinfo {author} {\bibfnamefont {C.~G.}\ \bibnamefont
  {Gray}},\ }\bibfield  {title} {\enquote {\bibinfo {title} {Definition of the
  magnetic quadrupole moment},}\ }\href {https://doi.org/10.1119/1.12213}
  {\bibfield  {journal} {\bibinfo  {journal} {American Journal of Physics}\
  }\textbf {\bibinfo {volume} {48}},\ \bibinfo {pages} {984--985} (\bibinfo
  {year} {1980})},\ \Eprint
  {https://arxiv.org/abs/https://doi.org/10.1119/1.12213}
  {https://doi.org/10.1119/1.12213} \BibitemShut {NoStop}%
\bibitem [{\citenamefont {GonzÃ¡lez}\ \emph {et~al.}(1998)\citenamefont
  {GonzÃ¡lez}, \citenamefont {JuÃ¡rez}, \citenamefont {Kielanowski},\ and\
  \citenamefont {Loewe}}]{StillScaryAJP}%
  \BibitemOpen
  \bibfield  {author} {\bibinfo {author} {\bibfnamefont {H.}~\bibnamefont
  {Gonz\'{a}lez}}, \bibinfo {author} {\bibfnamefont {S.~R.}\ \bibnamefont
  {Ju\'{a}rez}}, \bibinfo {author} {\bibfnamefont {P.}~\bibnamefont
  {Kielanowski}},\ and\ \bibinfo {author} {\bibfnamefont {M.}~\bibnamefont
  {Loewe}},\ }\bibfield  {title} {\enquote {\bibinfo {title} {Multipole
  expansion in magnetostatics},}\ }\href {https://doi.org/10.1119/1.18850}
  {\bibfield  {journal} {\bibinfo  {journal} {American Journal of Physics}\
  }\textbf {\bibinfo {volume} {66}},\ \bibinfo {pages} {228--231} (\bibinfo
  {year} {1998})},\ \Eprint
  {https://arxiv.org/abs/https://doi.org/10.1119/1.18850}
  {https://doi.org/10.1119/1.18850} \BibitemShut {NoStop}%
\bibitem [{CMS()}]{CMSmags}%
  \BibitemOpen
  \href@noop {} {}\bibinfo {howpublished} {\href{CMS
  Magnetics}{https://cmsmagnetics.com/}},\ \bibinfo {note} {accessed:
  2021-08-09}\BibitemShut {NoStop}%
\bibitem [{Sha()}]{Shapeways}%
  \BibitemOpen
  \href@noop {} {}\bibinfo {howpublished}
  {\href{Shapeways}{https://www.shapeways.com/}},\ \bibinfo {note} {accessed:
  2021-08-09}\BibitemShut {NoStop}%
\bibitem [{Phy()}]{Phyphox}%
  \BibitemOpen
  \href@noop {} {}\bibinfo {howpublished}
  {\href{Phyphox}{https://phyphox.org/}},\ \bibinfo {note} {accessed:
  2021-08-09}\BibitemShut {NoStop}%
\bibitem [{log()}]{loglog}%
  \BibitemOpen
  \href@noop {} {}\bibinfo {note} {Generally, care must be taken when using
  log-log plots, and the results are usually only valid over many decades of
  data.}\BibitemShut {Stop}%
\bibitem [{Note2()}]{Note2}%
  \BibitemOpen
  \bibinfo {note} {We are indebted to our editors and anonymous reviewers for
  many of these suggestions.}\BibitemShut {Stop}%
\end{thebibliography}

\begin{thebibliography}{4}%
\makeatletter
\addtocounter{NAT@ctr}{36}
\providecommand \@ifxundefined [1]{%
 \@ifx{#1\undefined}
}%
\providecommand \@ifnum [1]{%
 \ifnum #1\expandafter \@firstoftwo
 \else \expandafter \@secondoftwo
 \fi
}%
\providecommand \@ifx [1]{%
 \ifx #1\expandafter \@firstoftwo
 \else \expandafter \@secondoftwo
 \fi
}%
\providecommand \natexlab [1]{#1}%
\providecommand \enquote  [1]{``#1''}%
\providecommand \bibnamefont  [1]{#1}%
\providecommand \bibfnamefont [1]{#1}%
\providecommand \citenamefont [1]{#1}%
\providecommand \href@noop [0]{\@secondoftwo}%
\providecommand \href [0]{\begingroup \@sanitize@url \@href}%
\providecommand \@href[1]{\@@startlink{#1}\@@href}%
\providecommand \@@href[1]{\endgroup#1\@@endlink}%
\providecommand \@sanitize@url [0]{\catcode `\\12\catcode `\$12\catcode
  `\&12\catcode `\#12\catcode `\^12\catcode `\_12\catcode `\%12\relax}%
\providecommand \@@startlink[1]{}%
\providecommand \@@endlink[0]{}%
\providecommand \url  [0]{\begingroup\@sanitize@url \@url }%
\providecommand \@url [1]{\endgroup\@href {#1}{\urlprefix }}%
\providecommand \urlprefix  [0]{URL }%
\providecommand \Eprint [0]{\href }%
\providecommand \doibase [0]{https://doi.org/}%
\providecommand \selectlanguage [0]{\@gobble}%
\providecommand \bibinfo  [0]{\@secondoftwo}%
\providecommand \bibfield  [0]{\@secondoftwo}%
\providecommand \translation [1]{[#1]}%
\providecommand \BibitemOpen [0]{}%
\providecommand \bibitemStop [0]{}%
\providecommand \bibitemNoStop [0]{.\EOS\space}%
\providecommand \EOS [0]{\spacefactor3000\relax}%
\providecommand \BibitemShut  [1]{\csname bibitem#1\endcsname}%
\let\auto@bib@innerbib\@empty
\bibitem [{\citenamefont {Monteiro}\ \emph {et~al.}(2021)\citenamefont
  {Monteiro}, \citenamefont {Stari}, \citenamefont {Cabeza},\ and\
  \citenamefont {MartÃ­}}]{SmartphoneErrorAJP}%
  \BibitemOpen
  \bibfield  {author} {\bibinfo {author} {\bibfnamefont {M.}~\bibnamefont
  {Monteiro}}, \bibinfo {author} {\bibfnamefont {C.}~\bibnamefont {Stari}},
  \bibinfo {author} {\bibfnamefont {C.}~\bibnamefont {Cabeza}},\ and\ \bibinfo
  {author} {\bibfnamefont {A.~C.}\ \bibnamefont {Mart\'{i}}},\ }\bibfield  {title}
  {\enquote {\bibinfo {title} {Using mobile-device sensors to teach students
  error analysis},}\ }\href {https://doi.org/10.1119/10.0002906} {\bibfield
  {journal} {\bibinfo  {journal} {American Journal of Physics}\ }\textbf
  {\bibinfo {volume} {89}},\ \bibinfo {pages} {477--481} (\bibinfo {year}
  {2021})},\ \Eprint {https://arxiv.org/abs/https://doi.org/10.1119/10.0002906}
  {https://doi.org/10.1119/10.0002906} \BibitemShut {NoStop}%
\bibitem [{\citenamefont {Gray}\ and\ \citenamefont
  {Gubbins}(1984)}]{graybook}%
  \BibitemOpen
  \bibfield  {author} {\bibinfo {author} {\bibfnamefont {C.~G.}\ \bibnamefont
  {Gray}}\ and\ \bibinfo {author} {\bibfnamefont {K.~E.}\ \bibnamefont
  {Gubbins}},\ }\enquote {\bibinfo {title} {Theory of molecular fluids: I:
  Fundamentals},}\ in\ \href
  {https://doi.org/10.1093/oso/9780198556022.001.0001} {\emph {\bibinfo
  {booktitle} {Theory of Molecular Fluids: I: Fundamentals}}}\ (\bibinfo
  {publisher} {Oxford University Press},\ \bibinfo {year} {1984})\ pp.\
  \bibinfo {pages} {27--143},\ \bibinfo {edition} {1st}\ ed.\BibitemShut
  {Stop}%
\bibitem [{Note3()}]{Note3}%
  \BibitemOpen
  \bibinfo {note} {We are grateful to one of our anonymous reviewers who
  provided this derivation of magnetic fields outside of the multipole magnet
  configurations using vector calculus.}\BibitemShut {Stop}%
\bibitem [{\citenamefont {Jackson}(1999)}]{jackson}%
  \BibitemOpen
  \bibfield  {author} {\bibinfo {author} {\bibfnamefont {J.~D.}\ \bibnamefont
  {Jackson}},\ }\enquote {\bibinfo {title} {Classical electrodynamics},}\ in\
  \href@noop {} {\emph {\bibinfo {booktitle} {Classical Electrodynamics}}}\
  (\bibinfo  {publisher} {Oxford University Press},\ \bibinfo {year} {1999})\
  pp.\ \bibinfo {pages} {101--104},\ \bibinfo {edition} {3rd}\ ed.\BibitemShut
  {Stop}%
\end{thebibliography}

\providecommand{\noopsort}[1]{}\providecommand{\singleletter}[1]{#1}%
%



\pagebreak

\onecolumngrid
\begin{center}
  \textbf{\large Investigating the Magnetic Field Outside Small Accelerator Magnet Analogs via Experiment, Simulation, and Theory}

  \textbf{\large Supplementary Material}\\[.2cm]
\end{center}

\setcounter{equation}{0}
\setcounter{figure}{0}
\setcounter{table}{0}
\setcounter{section}{0}
\setcounter{page}{1}
\renewcommand{\theequation}{S\arabic{equation}}
\renewcommand{\thefigure}{S\arabic{figure}}
\renewcommand{\bibnumfmt}[1]{[S#1]}

\section{Experiment Setup and Calibration}

\begin{figure}
    \centering
    \includegraphics[width=0.7\linewidth]{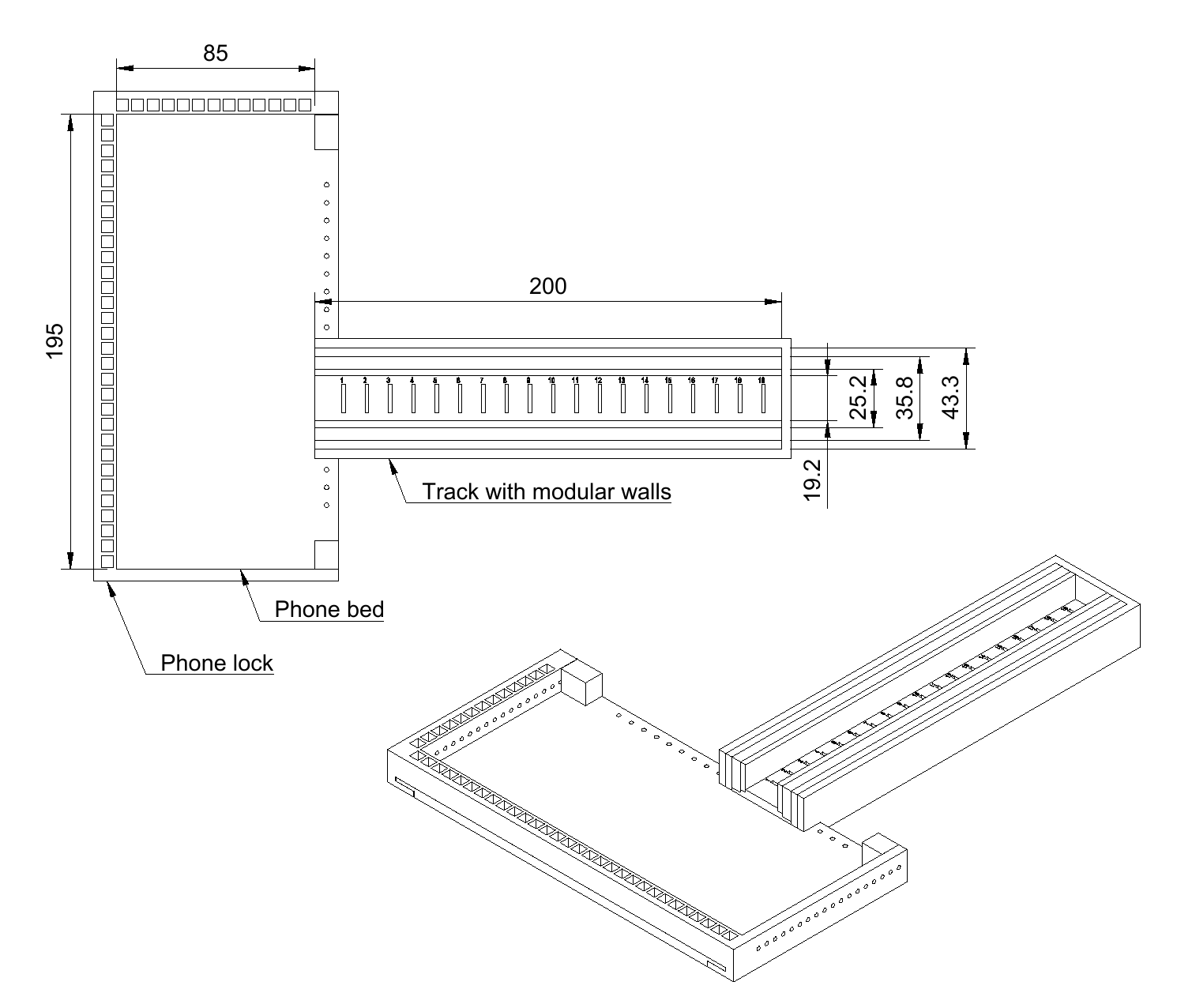}
    \includegraphics[width=0.7\linewidth]{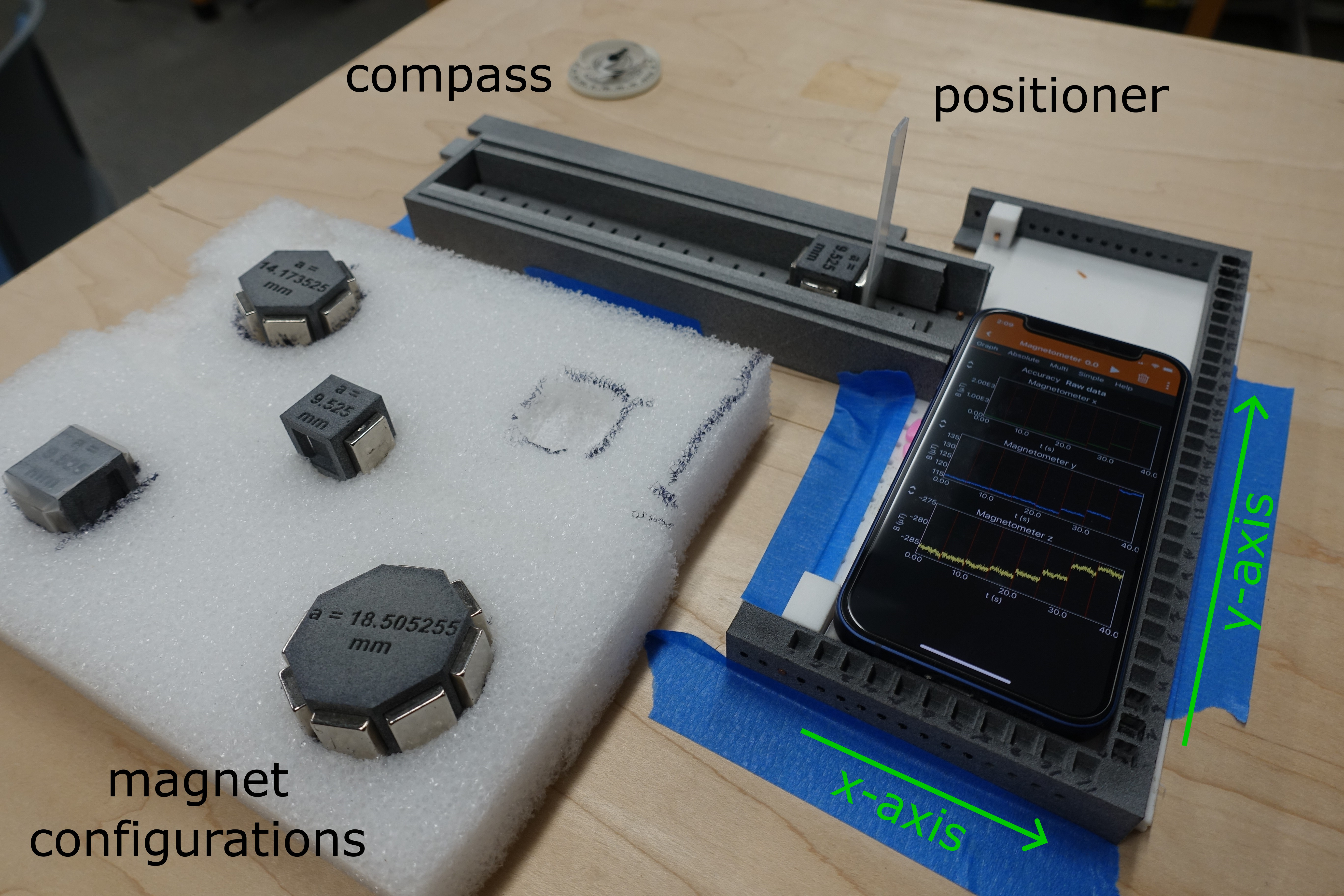}
    \caption{The experimental setup.  The upper image shows the CAD drawing of the experimental setup, including the smartphone bed and extension arm (or track), which can move relative to the smartphone bed.  All numbers have units of \si{mm}.  The lower image is the experimental setup in use, along with all the different magnet configurations.  The compass is used to align the setup to magnetic north (to eliminate Earth's magnetic field along the axis of measurement), the positioner rises from the extension arm to locate the magnet configuration along the arm.  The $x$- and $y$-axes of the phone are labeled.}
    \label{fig:CAD}
\end{figure}

In Fig.\ \ref{fig:CAD} we show both the CAD drawing drawing of the bed and arm and a photograph of the experiment.  The CAD files are available for download.\cite{EPAPS, supplementary}  In  Fig.\ \ref{fig:CAD}, we can also see the three different 3D-printed magnet holders (the quadrupole holder is also used to hold a single magnet for the dipole).  Care must be taken to keep the magnet configurations separate.  We ran our experiment on a piece of wood, raised by plastic above the surface of the table, in order to distance the setup from any metal screws in the table.

We used the app Phyphox\cite{Phyphox} to measure the magnetic field. Phyphox reports the magnetic field along all three axes of the phone, where the $x$-axis is along the short axis of the phone, the $y$-axis is along the long axis of the phone, and the $z$-axis points out of the front screen of the phone (forming a right-handed coordinate system).  It is important to note that the raw magnetic field measurement will include the Earth's field and any ambient field, including field from the phone itself and from local magnetized objects, such as steel table screws.  Students can reduce the effect of the Earth's field by aligning the apparatus with magnetic north. If perfectly aligned, this would remove the Earth's field from the direction of measurement entirely. For all measurements, the residual magnetic field measured without magnets present should be subtracted from the measurement.  Recognizing potential sources of systematic error in a measurement and carefully calibrating the apparatus and measurement tool to account for such errors is an important skill that students will hone in this experiment.

The holes along the edges of the smartphone bed walls in Fig.\ \ref{fig:CAD} are used to locate the sensor inside the smartphone.  The holes are designed to fit a cuboid $ 5 \times 5 \times 5$ \si{mm^3} alignment magnet at set locations along the phone.  The centers of the holes are separated by 6.5 mm.  Students can use the the simulation (Fig.\ \ref{fig: field contours}) to determine which magnetic pole should face the phone (usually north or south towards the phone works best).  Fig.\ \ref{fig: field contours} can also be used to predict the magnitude and sign of the magnetic field measured in the $x$ and $y$-directions as the alignment magnet is moved along each axis of the phone.  For the most accurate determination of the sensor location, the field should be measured along the $y$-axis using the holes on the short side of the phone and along the $x$-axis using the holes along the long side.  In both cases, students can predict that the magnitude of the measured field will be at a maximum when the magnet is aligned with the sensor and will drop off in value as the magnet moves away on either side of the sensor.

\begin{figure}
    \centering
    \includegraphics[width=0.47\linewidth]{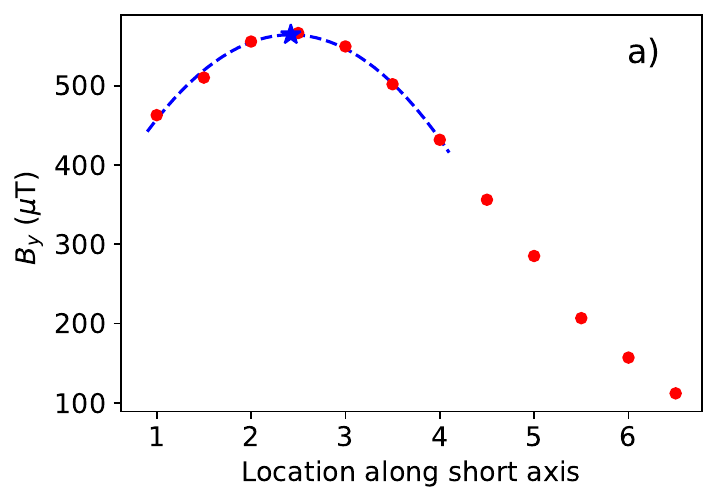}
    \includegraphics[width=0.47\linewidth]{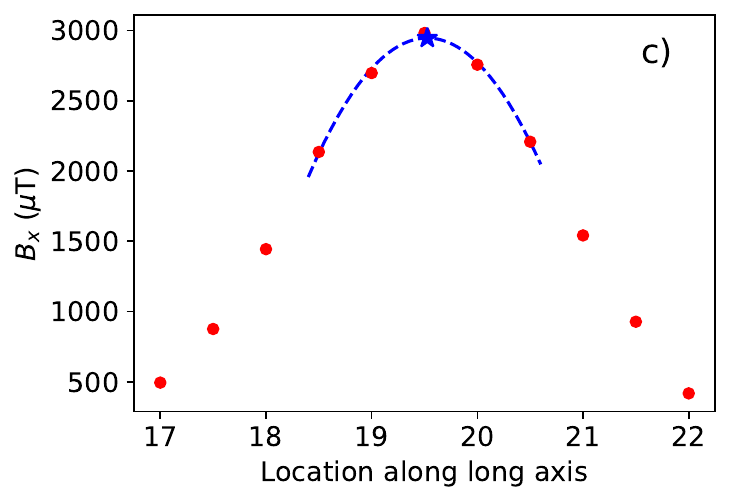}
    \includegraphics[width=0.47\linewidth]{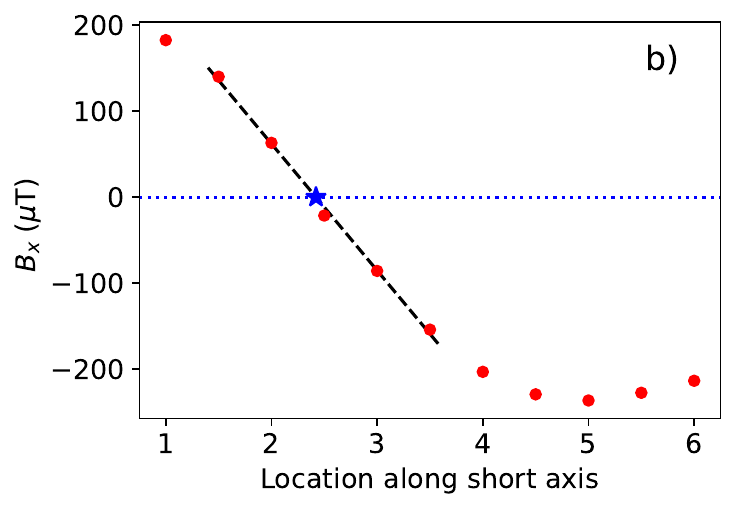}
    \includegraphics[width=0.47\linewidth]{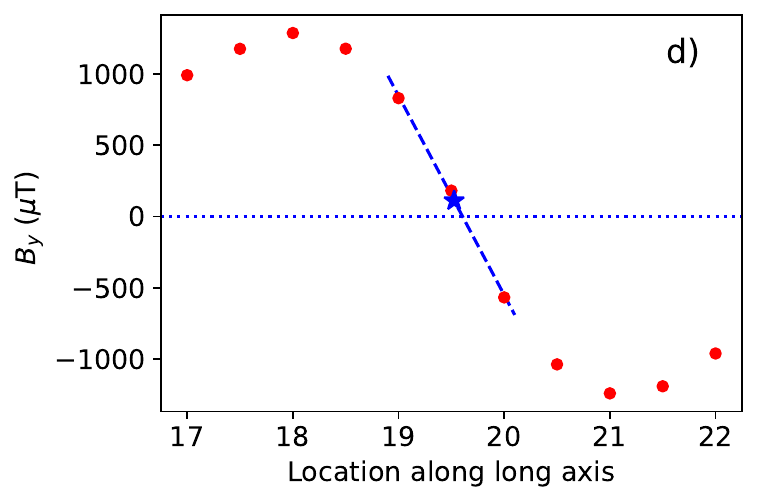}
    \caption{Determination of the location of the magnetic field sensor inside an iPhone 12 mini.  The positions correspond to the spaces for the small alignment magnet along the edge of the smartphone bed, and are separated by 6.5 mm.  (a) and (b) show the components of the field as a function of the magnet position along the short axis, (c) and (d) show the components of the field along the long axis.  The stars represent the location of the maximum of the quadratic fit (blue dashed line).  In (b) and (d), we find the sensor location where the linear fit crosses zero.  These locations are less than 0.5\% discrepant from the locations of the stars.}
    \label{fig:location}
\end{figure}

In order to increase our resolution, we held the alignment magnet just outside the holes and measured at the locations of the holes and in-between the holes as well.  A simple quadratic fit locates the maximum in the field, which corresponds to the location of the sensor along each axis.  The measurement tools in CAD can convert the position from locations relative to the holes to a position in real space (we used AutoDesk Fusion 360).  Example measurements of the location of the sensor in an iPhone12 mini are shown in Fig.\ \ref{fig:location}(a) and (c).  The uncertainty in the measurements\cite{SmartphoneErrorAJP} is smaller than the points.

For comparison, we also measured the field along the $x$-axis using the holes on the short side of the phone and along the $y$-axis using the holes along the long side (Fig.\ \ref{fig:location}(b) and (d)). In this case, students should predict that the direction of the field will switch sign as the alignment magnet passes across the location of the sensor and that the magnitude of the field will vary approximately linearly in the region close to the sensor. The precise location of the sensor is found by determining where a linear fit to the data crosses the zero of the magnetic field (when the non-zero offset in the absence of magnets is subtracted from the data). Further consideration will also reveal that the magnitude of the field farther from the sensor will reach a maximum then fall off again as the magnet continues to move away. Students must take care not to fit this peak, which places the sensor at a location far from the true location, leading to incorrect distances and results.

Determining the location of the sensor along the $z$-axis can be accomplished by moving a single permanent magnet along the extension arm toward and away from the phone and measuring the change in the $z$-component of the magnetic field.  After each measurement, the height of the phone in the bed is changed and the magnetic field is tested again.  To change the height, we used thin sheets of plastic or other non-magnetic (e.g., aluminum) material.  The magnetic field sensor is centered on the magnet when there is very little change in magnetic field along the $z$-axis as the magnet moves toward and away from the smartphone.

As a final step in preparing for data-taking, we measured the magnetic field strength of $23$ magnets at a distance of \SI{7.6}{cm} from the sensor and selected $8$ magnets with the smallest variation in magnetic field magnitude, thus also reducing variation in the dipole moments ($m$) of the permanent magnets used in the experiment. The complete collection of magnets exhibited a variation in $m$ of approximately $2.5\%$ while the smaller collection that was used for obtaining the data described in this paper varied by only $0.5\%$. Reducing the variation in the dipole moments of the individual magnets helps ensure that the lower-order moments will cancel for each magnet configuration, allowing the highest order term to dominate, as expected in our theoretical model (Eq.\ \ref{eq:generalfield}).  Preliminary data taken without considering the variation in the dipole moment of the permanent magnets yielded results significantly discrepant with our model for the sextupole and octupole magnet configurations (data not shown).

\section{Calculation of Multipole Moments}
The power-law dependence of the various multipoles is well-known\cite{griffiths_2017} and finding the power-law behavior alone can be a sufficient lab experience for many students.  The behavior of each higher-order multipole can be found by differentiating the previous multipole moment.  In this way, we see that $B\sim 1/r^{l+2}$, with $l=1,2,3,4$ for the dipole, quadrupole, octopole, and hexadecapole (16-pole) moments.  Log-log plots will show slopes of $-3$, $-4$, $-5$, and $-6$ for the dipole, quadrupole, sextupole, and octupole magnets.

Ambitious students will want to know what the intercept of the line means and how to derive it.  In order to fit both the slope \textbf{and} the intercept, we need a method to determine the precise field from a given multipole magnet configuration.  The derivation of the general multipole expansion (electric and magnetic) can be complex,\cite{HardMultipoleBook} but can be greatly simplified with rotational invariance arguments, using basic Cartesian and spherical harmonic tensor methods.\cite{gray09, *gray10, gray78, *HardMultipoleAJP, *gray79, *gray80, *StillScaryAJP,graybook}

In this Supplemental Material we present two methods of deriving the exact prefactors for the multipole magnet configurations.  The first method is appropriate for students in the upper-level who are familiar with vector calculus, the second (somewhat more tedious) method is appropriate for students at the introductory level, and requires only a knowledge of vectors.

Either of these methods can be used to verify the functional form presented in Eq.\ \ref{eq:generalfield} and the prefactors listed in Table \ref{tab:prefactors}.

\subsection{Calculation of multipole moments via vector calculus} \label{sec:vector}

We start with the magnetic field of a magnetic dipole $\vec{m}$:\footnote{We are grateful to one of our anonymous reviewers who provided this derivation of magnetic fields outside of the multipole magnet configurations using vector calculus.}
\begin{equation}
    \vec{B}_\mathrm{dip} = \frac{\mu_0 m}{4\pi r^3} (2 \cos(\theta) \hat{r} + \sin(\theta) \hat{\theta}),
    \label{Eq: dipole vector field again}
\end{equation}
where $\theta$ and $r$ are measured with respect to the location and direction of the dipole moment $\vec{m}$.  Away from this dipole, we can show that the curl of $\vec{B}$ is zero ($\nabla \times \vec{B} = 0$), which allows us to define a scalar potential for $\vec{B}$.  We recognize that the magnetic dipole has the same form as the electric field for an electric dipole, with $\mu_0 \vec{m}$ in place of $\vec{p}/\epsilon_0$.  We can use this fact to justify the scalar potential for a magnetic dipole:\cite{gray78, *HardMultipoleAJP, *gray79, *gray80, *StillScaryAJP}
\begin{equation}
    \Phi_\mathrm{dip} = \frac{\mu_0}{4\pi} \frac{\vec{m}\cdot\hat{r}}{r^2} = - \frac{\mu_0 }{4\pi}\, \vec{m}\cdot \nabla\frac{1}{r}
    \label{eq:dipolescalar}
\end{equation}

We wish to measure the field from the center of a collection of dipoles, each a distance $a$ from the center of the collection.  If you translate the dipole by $\vec{a}$, in the same direction as the dipole moment (such that $\hat{a} = \hat{m}$), the potential then becomes
\begin{equation}
    \Phi_{\mathrm{dip},\vec{a}} = - \frac{\mu_0 m}{4\pi}\, \hat{a}\cdot \nabla\left(\frac{1}{|\vec{r}-\vec{a}|}\right).
    \label{eq:dipolescalar2}
\end{equation}
The gradient in Eq.\ \ref{eq:dipolescalar2} is written with respect to $\vec{r}$.  Re-writing the gradient with respect to $\vec{a}$ gives
\begin{equation}
    \hat{a}\cdot \nabla\left(\frac{1}{|\vec{r}-\vec{a}|}\right) = - \hat{a}\cdot \nabla_{\vec{a}}\left(\frac{1}{|\vec{r}-\vec{a}|}\right) = -\frac{\partial}{\partial a}\left(\frac{1}{|\vec{r}-\vec{a}|}\right) .
    \label{eq:gradient}
\end{equation}
We can expand the term in parenthesis as follows:\cite{jackson}
\begin{equation}
    \left( \frac{1} {|\vec{r}-\vec{a}|} \right) = \sum_{l = 0}^\infty \frac{a^l}{r^{l+1}} P_l(\cos\gamma),
    \label{eq:sexpansion}
\end{equation}
where $P_l$ is the Legendre polynomial of order $l$ and $\gamma$ is the angle between $\vec{a}$ and $\vec{r}$.  Plugging Eq,\ \ref{eq:gradient} and \ref{eq:sexpansion} into Eq.\ \ref{eq:dipolescalar2}, we find:
\begin{equation}
    \Phi_{\mathrm{dip},\vec{a}} = \frac{\mu_0 m}{4\pi}\, \frac{\partial}{\partial a} \sum_{l = 0}^\infty\left[ \frac{a^l}{r^{l+1}} P_l(\cos\gamma)\right] = \frac{\mu_0 m}{4\pi}\, \sum_{l = 0}^\infty\left[ \frac{l\,a^{l-1}}{r^{l+1}} P_l(\cos\gamma)\right].
    \label{eq:sdipexpand}
\end{equation}

We can now use this scalar potential for a single dipole for all the dipoles in a given multipole configuration.  Suppose we have $N$ dipoles on the faces of a regular (planar) $N$-gon with polar coordinates of ($a$, $\theta_j$), where $\theta_j = 2\pi j/N$.  In order to make the multipole magnets shown in Fig.\ \ref{fig:multipoles from dipoles}, the dipole moment must alternate between pointing into the center and out from the center.  This means that $\vec{m}_j = (-1)^j\,m\, \hat{a}_j$.  Thus, the total scalar potential of all the dipoles on our planar $N$-gon would be:
\begin{equation}
    \Phi_{\mathrm{tot}} =  \frac{\mu_0 m}{4\pi}\, \sum_{l = 0}^\infty \sum_{j=0}^{N-1}  (-1)^j\left[ \frac{l\,a^{l-1}}{r^{l+1}} P_l(\cos(\theta - \theta_j))\right].
    \label{eq:spotential}
\end{equation}

Eq.\ \ref{eq:spotential} can be used to find the scalar potential at any value of $(r,\theta)$ outside of the multipole magnet configuration.  We have chosen to measure the magnetic field at a point which corresponds to the value of $\theta = 0$ (along the $x$-axis in Figs.\ \ref{fig:multipoles from dipoles}).  For $\theta = 0$, we have:
\begin{equation}
    \Phi_{\mathrm{tot}} =  \frac{\mu_0 m}{4\pi}\, \sum_{l = 0}^\infty  \frac{a^{l-1}}{r^{l+1}}\, g_l, \mathrm{~~~with}~~ g_l = l \sum_{j=0}^{N-1}  (-1)^j P_l(\cos\theta_j).
    \label{eq:spotential2}
\end{equation}

We find the magnetic field as the gradient of the scalar potential, $\vec{B} = - \nabla \Phi$.  With our choice to measure the field along $\theta = 0$, we see that there is only an $r$-component to the field along that axis.  Thus, the field becomes:
\begin{equation}
    B_r = -\frac{\partial \Phi}{\partial r} = \frac{\mu_0 m}{4\pi}\, \sum_{l = 0}^\infty  \frac{a^{l-1}}{r^{l+2}}\, f_l, \mathrm{~~~with}~~ f_l = l(l + 1)\,\sum_{j=0}^{N-1}  (-1)^j P_l(\cos\theta_j).
    \label{eq:sfinal}
\end{equation}

These sums can be evaluated either by hand or within software packages such as Mathematica or SageMath.  The first non-vanishing coefficient occurs at $l=N/2$.  Thus the leading term in the field for the planar quadrupole magnet ($N=4$) is the $l=2$, or quadrupole moment, with $f_2 = 18$. The leading term for the sextupole magnet ($N=6$) is $l=3$, the octopole moment, with $f_3 = 45$.  For the octupole magnet ($N=8$), the leading term in the magnetic field is $l=4$, the hexadecapole moment, with $f_4 = 175/2$.

Eqs.\ \ref{eq:spotential}, \ref{eq:spotential2}, and \ref{eq:sfinal} have been simplified for the case of dipoles on the faces of a planar $N$-gon.  For the linear quadrupole magnets, you can add two dipoles using Eq.\ \ref{eq:sdipexpand} to find:
\begin{eqnarray}
    \Phi_{\mathrm{tot}} &=& \pm \frac{\mu_0 m}{4\pi}\, \sum_{l = 0}^\infty \frac{l\,a^{l-1}}{r^{l+1}} \left[  P_l(\cos(\theta - 0^\circ)) + P_l(\cos(\theta - 180^\circ))\right]\\
    ~ &=&  \pm \frac{\mu_0 m}{4\pi}\, \sum_{l = 0}^\infty \frac{l\,a^{l-1}}{r^{l+1}} \left[  P_l(\cos\theta) + P_l(-\cos\theta)\right],
    \label{eq:spotentiallq}
\end{eqnarray}
where the two dipoles are at angles $0^\circ$ and $180^\circ$ degrees, and the $+$ sign is if the dipoles point away from the center, and the $-$ sign is if the dipoles point towards the center.  Because the Legendre polynomials are either even or odd depending on $l$, the term in hard brackets will always be zero for odd $l$ values.  Thus, the first non-zero term will be $l=2$, the quadrupole moment.  Then for the linear quadrupole magnets we have (to leading order):
\begin{equation}
    \Phi_{\mathrm{tot}} \approx  \pm \frac{\mu_0 m}{4\pi}\, \frac{2\,a}{r^{3}} \left[  3(\cos\theta)^2 -1\right].
    \label{eq:spotentiallqfin}
\end{equation}

We can use this to find the scalar potential of the linear quadrupole magnets when parallel to the track, which corresponds to $\theta = 0$ and the $+$ sign (dipole moments point out), and when the linear quadrupole magnets are perpendicular to the track, $\theta = 90$ and the $-$ sign (dipole moments point in).  We find:
\begin{equation}
    \Phi_{\mathrm{lin quad},\parallel} = \frac{\mu_0 m}{4\pi}\,\frac{4a}{r^3} ~~\mathrm{and}~~\Phi_{\mathrm{lin quad},\perp} = \frac{\mu_0 m}{4\pi}\,\frac{2a}{r^3}.
\end{equation}
In both cases, the field is only in the $r$-direction, and $B_r = -\partial \Phi/\partial r$ will give the values of $f$ found in Table \ref{tab:prefactors} of $f=12$ and $f=6$ for the linear quadrupole magnets parallel and perpendicular to the axis, respectively.

The power of this method is that it is relatively easy to evaluate the exact prefactors from Eqs.\ \ref{eq:sfinal} and \ref{eq:spotentiallqfin} for other angles.  Angles of interest might be along the line connecting the center to a vertex ($\theta =0$ is along the line that connects the center to a face).  Changing the angle can give different prefactors that can also be tested experimentally.

\subsection{Calculation of multipole moments via Taylor expansions} \label{sec:taylor}

If we look along the axis of the dipole, then $\theta=\SI{0}{\degree}$ in Eq.~\ref{Eq: dipole vector field again}.  Then the only field is the radial component, which for a dipole a distance $r$ from the center of a magnet is
\begin{equation}\label{eq:dipoleformula}
    B_{\mathrm{dip}} = \frac{\mu_0 m}{2\pi r^3}
\end{equation}
where $m$ is the dipole moment.\\

A linear quadrupole is constructed with two opposing dipoles separated by a distance $2a$.  Along the axis of the dipoles, we can use Eq.\ \ref{eq:dipoleformula} for each dipole to find the magnetic field a distance $r$ from the center of the configuration:

\begin{align*}
    B_{\mathrm{lin quad},\parallel} &= \frac{\mu_0 m}{2\pi} \left(\frac{1}{(r-a)^3} - \frac{1}{(r+a)^3}\right)\\
    ~  &= \frac{\mu_0 m}{2\pi r^3}\left(\left(1-\frac{a}{r}\right)^{-3} - \left(1+\frac{a}{r}\right)^{-3}\right)
\end{align*}
This can be simplified using the binomial approximation to find
\begin{equation}
    B_{\mathrm{lin quad},\parallel} \approx \frac{6\mu_0 ma}{2\pi r^4}
    \label{Eq:lQuadPar}
\end{equation}
\noindent in the region where $r \gg a$.

\begin{figure}
    \centering
    \includegraphics[width=0.9\linewidth]{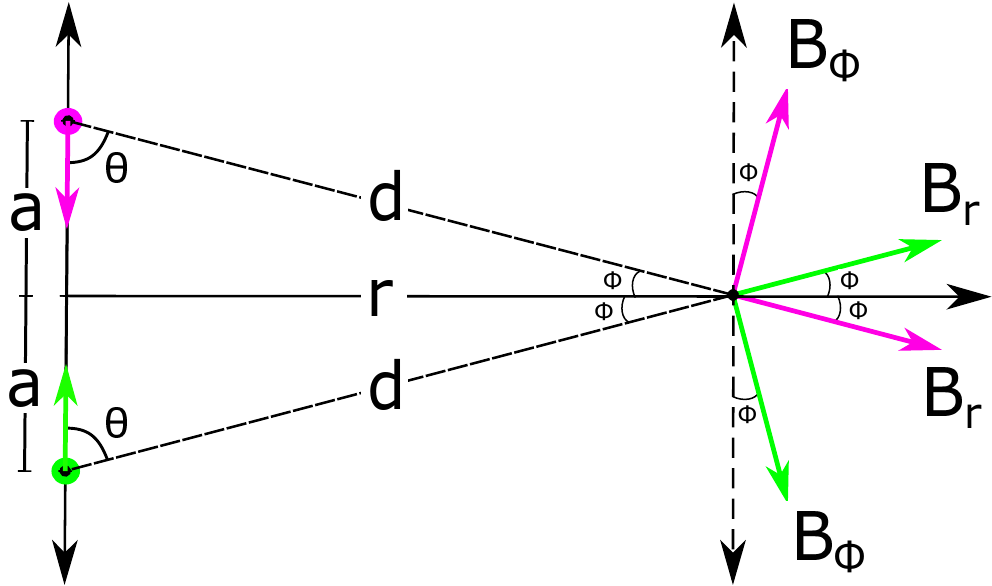}
    \caption{Calculation of the field from a linear quadrupole magnet oriented perpendicular to the measurement axis.  Each dipole creates a magnetic field which can be broken into components.  Only the component along the axis remains.}
    \label{Fig: calculations}
\end{figure}

A similar expression can be found along the axis that is perpendicular to the line connecting the two opposing dipoles.  If we again consider a point a distance $r$ from the center of the configuration, we see that each dipole creates fields in the $r$ and $\theta$ directions \textit{relative to the axis of that dipole}, meaning we must retain both terms from Eq.\ \ref{Eq: dipole vector field again}.  Each of those fields can be broken into components along the axis of interest and perpendicular to that axis, as shown in Fig.\ \ref{Fig: calculations}.

Due to symmetry, the fields perpendicular to the axis cancel each other out, leaving only a component along the axis from the center of the configuration to the point of interest.  Calling this field $B_{\mathrm{lin quad},\perp}$, and using the angles labeled in Fig.\ \ref{Fig: calculations}, we find:
\begin{align*}
    B_{\mathrm{lin quad},\perp} &= 2B_{\mathrm{dip}, r}\cos\phi + 2B_{\mathrm{dip}, \theta}\sin\phi\\
    ~ &= \frac{\mu_0 m}{2\pi d^3}\left(2\cos(\theta)\cos(\phi) + \sin(\theta)\sin(\phi)\right)\\
    ~ &= \frac{\mu_0 m}{2\pi d^3}\left(2\frac{a}{d}\cdot\frac{r}{d} + \frac{r}{d}\cdot\frac{a}{d}\right)\\
    ~ &= \frac{3\mu_0 ma}{2\pi d^5}r.
\end{align*}

Using $d^2 = r^2 + a^2$, we can modify the equation to be:
\begin{align*}
    B_{\mathrm{lin quad},\perp} &= \frac{3\mu_0 ma}{2\pi}r(r^2 + a^2)^{-5/2}\\
    ~ &= \frac{3\mu_0 ma}{2\pi r^4}\left(1 + \frac{a^2}{r^2}\right)^{-5/2}.
\end{align*}

This expression is exact but can be simplified if $r \gg a$. To first order in the binomial approximation, we find:
\begin{equation}
    B_{\mathrm{lin quad},\perp} \approx \frac{3\mu_0 ma}{2\pi r^4}
    \label{Eq:lQuadPerp}
\end{equation}

A planar quadrupole the is addition of two linear quadrupoles.  Adding Eq.~\ref{Eq:lQuadPar} and Eq.~\ref{Eq:lQuadPerp}, we find that the field of a planar quadrupole a distance $r$ from its center is
\begin{equation}
    B_{\mathrm{planar quad}} \approx \frac{9\mu_0 ma}{2\pi r^4}.
\end{equation}

A similar process of pairing dipoles, breaking into components, and expanding the resultant field magnitudes can be used to find the prefactors for the sextupole magnet and octupole magnet reported in Table \ref{tab:prefactors}.

\newpage

\providecommand{\noopsort}[1]{}\providecommand{\singleletter}[1]{#1}%

\end{document}